\documentclass[useAMS,usenatbib]{mn2e}

\usepackage{psfig, epsf, epsfig}

\title[LMC-Galaxy interaction]{The influences of the Magellanic Clouds
on the Galaxy: Pole shift, warp, and star formation history}
\author[K. Bekki]
{Kenji Bekki${}^1$\thanks{E-mail:
bekki@cyllene.uwa.edu.au} \\
${}^1$ICRAR M468
The University of Western Australia
35 Stirling Hwy, Crawley
Western Australia, 6009}

\begin{document}

\date{Accepted, Received 2005 February 20; in original form }

\pagerange{\pageref{firstpage}--\pageref{lastpage}} \pubyear{2005}

\maketitle

\label{firstpage}

\begin{abstract}
We investigate
how the Large Magellanic Cloud (LMC) influences the evolution of
the Galaxy after the LMC enters into the virial radius of the dark matter
halo of the Galaxy for the first time. Both the Galaxy and the LMC
are modeled as N-body particles in our models
so that the dynamical influences of the LMC on the Galaxy
can be investigated in a fully self-consistent manner.
Furthermore, the orbital parameters for the LMC are carefully
chosen such that the present location of the LMC in the Galaxy
can be rather
precisely reproduced in our simulations.  We particularly
investigate the influences of the LMC on the precession rate,
the outer stellar and gaseous structures,  
and the star formation history of the Galaxy.
Our principals results are summarized as follows.
The LMC-Galaxy dynamical interaction can cause  
``pole shift'' (or irregular precession/nutation)
of the Galaxy and  the typical rate of pole shift  ($\dot{ {\theta}_{\rm d} }$)
is $\sim 2$ degree Gyr$^{-1}$ corresponding to
$\sim 7 \mu$as yr$^{-1}$.
The LMC-Galaxy interaction induces the formation of  the outer warp structures 
of the Galaxy,
which thus confirms the results of previous numerical simulations
on the formation of the Galactic warp. 
The LMC-Galaxy interaction also induces the formation of outer
($R>20$ kpc) spiral arms and increases the vertical velocity dispersion
of the outer disk significantly.
The mean star formation rate of the Galaxy for the last
several Gyrs can be hardly influenced by the LMC's tidal force.
The age and metallicity distribution of stars  in the solar-neighborhood
(7 kpc $\le R\le $ 10 kpc) for the last several Gyr
can be only slightly changed by the past LMC-Galaxy interaction.
If the LMC is accreted
onto the Galaxy as a group with small dwarf galaxies,
then the stripped dwarfs forms a unique polar  distribution 
within the Galaxy.
Based on these results, we discuss how the possible ongoing Galactic
pole shift  with $\dot{ {\theta}_{\rm d} } \sim 10$ $\mu$as yr$^{-1}$ 
can be detected by future observational studies by GAIA.
\end{abstract}


\section{Introduction}

The tidal field of the Galaxy and the hydrodynamical influences of
the warm gaseous halo have long been considered to be key ingredients
in the evolution of the LMC and the Small Magellanic Cloud (SMC).
The strong gravitational field of the massive dark matter halo
of the Galaxy can influence not only the dynamical evolution
of the LMC (e.g., Gardiner et al. 1994) but also its long-term  star formation
history (Bekki \& Chiba 2005; BC05).
The hydrodynamical interaction between the Galactic halo and cold interstellar
medium (ISM) of the LMC can be responsible for the stripping of ISM from
the LMC (e.g., Mastropietro et al. 2005; M05). The strong tidal field
of the Galaxy  can play
a vital role in the formation of the bifurcated structures of the Magellanic
Stream and its elongated leading arm 
features (e.g., Gardiner \& Noguchi 1996, GN96;
Yoshizawa \& Noguchi 2003, YN03; Connors et al. 2006; 
Diaz \& Bekki 2011a, b, c, DB11a, b, and c, respectively). 
The orbital evolution of the LMC and the SMC,
which is essentially important for the evolution of the Clouds,
strongly depend on the mass model of the Galaxy and the present
day proper motions of the Clouds (e.g., Lin \& Lynden-Bell 1977;
Murai \& Fujimoto 1980; Kallivayalil et al. 2006; Besla et al. 2007, B07;
Ruzicka et al. 2010).

Although previous theoretical studies clarified 
the important influences of the LMC-Galaxy interaction
on the evolution of the LMC (e.g., BC05; M05),
their did not discuss how the LMC-Galaxy interaction influences
the evolution of the Galaxy.
Furthermore the influences of the LMC on the Galaxy have been so far
discussed only in the context of  the formation processes of the
Galactic warps (e.g., Hunter \& Toomre 1969; Weinberg 1998; Tsuchiya 2002).
Previous observations on the long-term star formation history in
the solar-neighborhood suggested that the star formation history
(in particular, a number of possible bursts of star formation)  might
have been influenced by the LMC-Galaxy tidal interaction (e.g., Rocha-Pinto
et al. 2000). The possible influences of the LMC on the Galactic star 
formation history, however,
have not been discussed so far by theoretical studies  
based on numerical simulations. The previous numerical simulations 
on the Galactic warp formation  did not consider the latest
observational results on  the proper motions of the LMC and the SMC
(e.g., Costa et al. 2009; Vieira et al. 2010).
Thus more self-consistent numerical simulations are necessary to address
a number of key questions related to the Galaxy evolution influenced 
by the LMC.

The purpose of this paper is thus to investigate
the influences of the LMC-Galaxy interaction on the  evolution of the Galaxy
using more realistic and more self-consistent numerical simulations.
The present study is significantly improved over the previous theoretical
studies as follows. First, the Galaxy is modeled  as a three-component
system (bulge, disk, and dark matter halo) represented by N-body particles 
so that both dynamical friction of the LMC against the Galactic halo
and the structural and kinematical changes of the Galactic halo due to
the LMC-Galaxy interaction can be self-consistently investigated.
Second, not only the orbital
evolution of the LMC since its first infall onto the Galaxy from outside  the 
virial radius but also  the mass evolution (e.g., mass loss)  of the LMC
are investigated so that the long-term influences of the LMC on the Galaxy
can be quantitatively estimated. 
Furthermore, star formation processes and gas dynamics in the  Galaxy
are included in some models of the present study
(though they are idealized in 
some points) so that the influences of the LMC on the Galactic star formation
history can be investigated.

We consider that one of the important influences of the LMC on the Galaxy
is the induction of the Galactic precession/nutation 
(or ``pole shift'') in the present study. 
As discussed later in this paper,
if the precession of the Galaxy is really ongoing,
it has a number of  potentially  important implications 
in the modern astronomy. 
Thus it is doubtless worthwhile for the present study to predict the possible
direction and rate  of the Galactic precession/nutation/pole-shift  
($\dot{ {\theta}_{\rm d} }$) caused  by the past LMC-Galaxy interaction
and thereby discuss whether the precession/nutation/pole-shift
 can be detected by future 
observational studies by GAIA and VLBI.

It is essential for the present study to reproduce rather precisely
the present location of the LMC with respect to the Galactic center,
because we discuss the present structure and kinematics of the Galaxy 
influenced by the LMC. Previous studies on the past orbital evolution
of the LMC (e.g., MF80; YN03) are based on the ``backward integration scheme''
in which the past orbit of the LMC can be readily investigated
for given present 
three dimensional (3D) position and velocity of the LMC.
In the present study,  we do not adopt the backward integration
scheme but instead investigate the orbital evolution of the LMC
from the past to the present by using N-body simulations. 
Therefore, we need to run at least a number of models with different initial
locations and velocities of the LMC in order to find a model
which can reproduce the present location of the LMC. 
It is accordingly numerically costly for the present study
to find a model that reproduces well the present location of the LMC.
But the present numerical simulations need to be done to properly investigate 
the influences of the Magellanic Clouds on the Galaxy evolution.

The layout of this paper is as follows. In \S 2, we summarize our numerical
models for the LMC-Galaxy interaction and describe the method to find
a model that reproduces the present location of the LMC. 
We present the 
results of the collisionless simulations
on the Galactic structure and kinematics influenced
by the LMC  in \S 3.
In \S 4, we briefly discuss how the Galactic star formation history can be
influenced by the LMC-Galaxy interaction using models with gas dynamics
and star formation.
In this section,
we also present some important implications
of the present numerical results in the long-term evolution
of the Galaxy.
The conclusions of the present study are given in \S 5.

\begin{table*}
\centering
\begin{minipage}{175mm}
\caption{Description of the  model parameters for
the representative models.}
\begin{tabular}{cccccccccc}
{Model}
& {$M_{\rm dm, mw}$
\footnote{The total dark halo mass 
of the Galaxy  in units of ${\rm M}_{\odot}$.}}
& {$r_{\rm vir, mw}$
\footnote{The virial radius of the Galactic dark matter halo  in units of kpc.}}
& {$M_{\rm dm, l}$
\footnote{The total dark halo mass 
of the LMC  in units of ${\rm M}_{\odot}$.}}
& {$R_{\rm edc}$
\footnote{The size of the extended disk component (EDC)
in the Galaxy in units of kpc}}
& {$f_{\rm v}$
\footnote{The ratio of the velocity of the LMC ($v_{\rm L}$)
to the circular velocity  ($v_{\rm cir}$)
at the initial position of the LMC .}}
& {$f_{\rm v, t}$
\footnote{The initial tangential  velocity of the LMC is given as $f_{\rm v,t}v_{\rm cir}$.}}
& {$f_{\rm v, r}$
\footnote{The initial radial  velocity of the LMC is given as $f_{\rm v,r}v_{\rm cir}$.}}
& {Dwarfs 
\footnote{Presence (``Yes'') or absence (``No'') of dwarf 
galaxies of the LMC group  in the model.}}
& {comments }\\
T1 & $10^{12}$  & 245 &  $9 \times 10^{10}$ & 35 & 0.63 & 0.6 & 0.2 &  
No & the standard model \\
T2 & $10^{12}$  & 245 &  $9 \times 10^{10}$ & 35 & 0.45 & 0.4 & 0.2 &  
No & \\
T3 & $10^{12}$  & 245 &  $9 \times 10^{10}$ & 35 & 0.54 & 0.5 & 0.2 &  
No & \\
T4 & $10^{12}$  & 245 &  $9 \times 10^{10}$ & 35 & 0.73 & 0.7 & 0.2 &  
No & \\
T5 & $10^{12}$  & 245 &  $9 \times 10^{10}$ & 35 & 0.60 & 0.6 & 0.0 &  
No & \\
T6 & $10^{12}$  & 245 &  $9 \times 10^{10}$ & 35 & 0.85 & 0.6 & 0.6 &  
No & \\
T7 & $10^{12}$  & 245 &  $9 \times 10^{10}$ & 35 & 1.17 & 0.6 & 1.0 &  
No & \\
T8 & $10^{12}$  & 245 &  $9 \times 10^{10}$ & 35 & 1.35 & 0.9 & 1.0 &  
No & \\
T9 & $10^{12}$  & 245 &  $3 \times 10^{10}$ & 35 & 0.63 & 0.6 & 0.2 &  
No & low-mass LMC\\
T10 & $10^{12}$  & 245 &  $6 \times 10^{10}$ & 35 & 0.63 & 0.6 & 0.2 &  
No & \\
T11 & $10^{12}$  & 245 &  $1.2 \times 10^{11}$ & 35 & 0.63 & 0.6 & 0.2 &  
No & high-mass LMC\\
T12 & $10^{12}$  & 245 &  $3 \times 10^{10}$ & 35 & 0.45 & 0.4 & 0.2 &  
No & \\
T13 & $10^{12}$  & 245 &  $3 \times 10^{10}$ & 35 & 0.54 & 0.5 & 0.2 &  
No & \\
T14 & $10^{12}$  & 245 &  $6 \times 10^{10}$ & 35 & 0.63 & 0.7 & 0.2 &  
No & \\
T16 & $10^{12}$  & 245 &  $9 \times 10^{10}$ & 25
& 0.63 & 0.6 & 0.2 &  
No & smaller EDC \\
T17 & $10^{12}$  & 245 &  $9 \times 10^{10}$ & 45
& 0.63 & 0.6 & 0.2 &  
No & larger EDC \\
T18 & $10^{12}$  & 245 &  $9 \times 10^{10}$ & 35
& 1.12 & 0.5 & 1.0 &  
No & first passage model\\
T19 & $10^{12}$  & 245 &  $3 \times 10^{10}$ & 35
& 1.12 & 0.5 & 1.0 & 
No & \\
T20 & $10^{12}$  & 245 &  $1.2 \times 10^{11}$ & 35
& 1.12 & 0.5 & 1.0 &  
No & \\
T21 & $5 \times 10^{11}$  & 245 &  $3 \times 10^{10}$ & 35 
& 0.63 & 0.6 & 0.2 &  
No & low-mass Galaxy\\
T22 & $5 \times 10^{11}$  & 245 &  $9 \times 10^{10}$ & 35 
& 0.63 & 0.6 & 0.2 &  
No & \\
T23 & $10^{12}$  & 245 &  $9 \times 10^{10}$ & 25
& 0.63 & 0.6 & 0.2 &  
Yes &  20 dwarfs, $R_{\rm dw}=10R_{\rm d,l}$ \\
T24 & $10^{12}$  & 245 &  $9 \times 10^{10}$ & 25
& 0.63 & 0.6 & 0.2 &  
Yes &  20 dwarfs, $R_{\rm dw}=5R_{\rm d,l}$ \\
I1 & $10^{12}$  & 245 &  $-$ & 35 & - & - & - &  
No & isolated model\\
I2 & $10^{12}$  & 245 &  $-$ & 25 & - & - & - &  
No & \\
I3 & $10^{12}$  & 245 &  $-$ & 45 & - & - & - &  
No & \\
\end{tabular}
\end{minipage}
\end{table*}

\section{The collisionless model}

\subsection{The Galaxy}

Since the present model of the Galaxy is essentially
the same as that adopted in our previous study
on the dynamical evolution of disk galaxies (Bekki \& Peng 2006),
we here briefly describe the model.
A much larger number of particles ($>10^6$) are used for a galaxy
in the present study (in comparison with our previous work above)
so that we can discuss evolution of global disk structures under the
influences of weaker tidal perturbation.  
The total  mass and the virial radius  of the dark matter halo  of the galaxy 
are denoted as $M_{\rm dm, mw}$ and $r_{\rm vir, mw}$, respectively.
We adopted an NFW halo density distribution (Navarro, Frenk \& White 1996)
suggested from CDM simulations:
\begin{equation}
{\rho}(r)=\frac{\rho_{0}}{(r/r_{\rm s})(1+r/r_{\rm s})^2},
\end{equation}
where  $r$, $\rho_{0}$, and $r_{\rm s}$ are
the spherical radius,  the characteristic  density of a dark halo,  and the
scale
length of the halo, respectively.
We adopted $c=10$ ($=r_{\rm vir, mw}/r_{\rm s}$)
and $r_{\rm vir, mw}=245$ kpc, 
and the mass ratio of halo to disk
was fixed at 16.7 ($M_{\rm dm, mw}=10^{12} M_{\odot}$
for the adopted disk mass). The adopted total mass of the Galaxy
is consistent with the observationally estimated value of
$1.9^{+3.6}_{-1.7} \times 10^{12} {\rm M}_{\odot}$
(Wilkinson \& Evans 1999).

The stellar disk with the size $R_{\rm d}$ 
and the mass  $M_{\rm d}$
is assumed to have an exponential
profile with the radial and vertical scale lengths of 3.5 kpc and 0.35 kpc,
respectively.
The stellar disk is assumed to have  $M_{\rm d}=6 \times 10^{10} {\rm M}_{\odot}$,
$R_{\rm d}=17.5$ kpc, the Toomre Q-parameter of 1.5,
and  the maximum circular velocity of  $\sim 220$ km s$^{-1}$.
The bulge is represented by a Hernquist bulge
and has a mass of $10^{10} {\rm M}_{\odot}$ and a size of 3.5 kpc
(scale-length of 700pc). 
The disk galaxy is assumed to have no gas
and no star formation. The initial disk plane of the galaxy is set to be the $x$-$y$ plane
in the present study (i.e., the $z$-axis is the polar direction of the galaxy).
Fig. 1 shows the radial profile of the circular 
velocity ($V_{\rm c}$) of the present
disk model.

Recent observational studies have revealed the detailed structural 
properties of the  outer stellar and gaseous disks of the Galaxy 
(e.g., McClure-Griffiths, et al. 2004;
Levine et al. 2006, 2007;  Momany et al. 2006).
In order to investigate the dynamical influences of 
the LMC on the outer
part ($R>20$ kpc) of the Galactic stellar and gaseous disks,
we assume that the Galaxy has an extended disk component (referred to
as ``EDC'' from now on for convenience).
Although  the present simulations are purely collisionless
so that  we can not
investigate the LMC's influences on the outer gas disk of the Galaxy,
the inclusion of the EDC in the present simulations can help us to
understand the LMC's influences on the outer part of the Galaxy.
The mass fraction of the EDC to the Galactic stellar disk
is fixed at 0.1 
(i.e., the EDC's mass of $6 \times 10^9 {\rm M}_{\odot}$) and the size
($R_{\rm edc}$) is a free parameter ranging from 25 kpc to 45 kpc.
The radial scale length of the EDC is set to be $0.5R_{\rm edc}$
for all models whereas two different values for
the vertical scale height
($0.0025R_{\rm edc}$ and $0.01R_{\rm edc}$) are adopted.

The total numbers of  particles used for dark matter halo, stellar disk,
bulge, and EDC  of a disk galaxy in a simulation are 800000, 200000, 33400, and 
50000, respectively.
This simulation with these particle numbers is referred to as 
a ``high-resolution'' simulation for convenience, just
because the particle numbers are larger than those used in ``low-resolution
simulations later described.
The adopted  gravitational softening
length is fixed at 200pc for all components
(e.g., dark matter halo and disk) in all high-resolution  models.
We investigate some models with the total particle number of EDC being
200000 in order to confirm that the present results
(in particular, the formation of the Galactic warp)  do not depend
on the resolution of the simulations.
Table 2 briefly summarizes the particle mass and the total particle
number of each component used in the present simulations.

We investigate models in which the Galaxy is not influenced
by the LMC at all and they are referred to as ``isolated models'' 
(e.g., I1). In order to confirm that the central bar of the Galaxy
can be formed in the isolated model,
we investigate the dynamical evolution of the isolated model.
Fig. 2 shows the final two-dimensional (2D) distribution of disk stars 
projected onto the $x$-$y$ plane
after 5.2 Gyr dynamical evolution of  the isolated
Galaxy model (I1). The 2D surface mass density 
($\mu_{\rm s}$) is 
given in logarithmic scale in this figure and all other
figures on the 2D distributions of stars.
The disk clearly
has a central bar/box bulge, which is consistent with the observed
Galaxy. It should be noted here that
the global stellar bar of the Galaxy  can be formed {\it spontaneously}
as a result of bar instability without tidal interaction with the LMC. 
It is confirmed that the EDC does not show any warps in the final disk
for this isolated model.

The mass of the Galaxy at the epoch of the LMC accretion could be 
significantly smaller than the present mass of the Galaxy owing to 
the continuous growth via accretion of gas, dark matter, and small galaxies.
We therefore investigate the ``low-mass'' model for the Galaxy in which
$M_{\rm dm, mw}$ and $r_{\rm vir, mw}$ are $5 \times 10^{11} {\rm M}_{\odot}$
and 206 kpc, respectively. The Galaxy can be more strongly influenced
by the LMC for a given mass of the LMC in this low-mass Galaxy model.
The total mass of the Galaxy is fixed for all models in the present study.
It is our future study to investigate the orbital evolution of the LMC
in the time-changing gravitational potential due to the mass growth of
the Galaxy.

\subsection{The LMC}

The LMC is assumed to be accreted onto the Galaxy from outside the virial
radius of the dark matter halo ($R>245$ kpc). Therefore, the LMC
has a dark matter halo that is not tidally truncated and thus
can have a total mass  much larger
than that of the present dark matter halo of the LMC
($\sim 10^{10} {\rm M}_{\odot}$).
The dark matter halo of the LMC is assumed to have
the NFW profile with $c=12$, and the total mass of the halo ($M_{\rm dm, l}$) 
within the virial radius ($r_{\rm vir, l}$) 
is a free parameter that controls 
the orbital evolution of the LMC.
The LMC is assumed to be a bulge-less dwarf disk galaxy 
with the disk mass $M_{\rm d,l}$
and the disk size $R_{\rm d,l}$.
The mass-ratio of the disk to the halo is fixed at 16.7 for all models
in the present study.
The LMC has an exponential disk with
the scale length and vertical scale-hight are fixed at
$0.2R_{\rm d,l}$ and and $0.02R_{\rm d, l}$, respectively.
The Toomre Q-parameter of the disk is set to be 1.5 for all models.

We mainly investigate the LMC models with 
$M_{\rm dm, l}=9 \times 10^{10} {\rm M}_{\odot}$,
$M_{\rm d,l}=5.4 \times  10^9 {\rm M}_{\odot}$,
and $r_{\rm vir, l}=74.9$ kpc. This model shows the maximum circular
velocity ($v_{\rm max,l}$)
of $\sim 110$ km s$^{-1}$, which is consistent with
some observational results (e.g., Olsen \& Massey 2007;
Piatek et al. 2008).  We however investigate 
models with different $M_{\rm dm, l}$, because  observational
studies of $v_{\rm max,l}$  derived from kinematic of different
stellar populations show different results
(e.g., Olsen \& Massey 2007).
As suggested by our previous study (Bekki 2008), the LMC 
was accreted onto the Galaxy as a group with some satellite dwarf galaxies.
Other faint group member galaxies are not distributed
within the LMC halo from most models  in the present study,
because they can not be so important in the orbital evolution of the LMC
(group) owing to the much smaller  total mass of the dwarfs
in comparison with the dark matter of the LMC.

We however investigate several models with 20 small dwarfs in order
to discuss the present distribution of the stripped dwarfs in the 
outer region of the Galactic halo.
The masses of the 20 dwarfs are assumed to be the same
($0.001 M_{\rm d, l}$ for each) and they are distributed
within  a radius of $R_{\rm dw}$. The radial number distribution
of the dwarfs is assumed to follow the power-law profile
with the power-law index of $-3.5$ and the scale length of the
profile of $0.5R_{\rm dw}$. We investigate models with
different $R_{\rm dw}$ so that we can discuss how the present
distribution of the stripped dwarfs depends on the initial distribution
of the dwarfs in the LMC. The results are discussed in \S4.

Here we do not include the influences of the SMC on the orbit of the LMC
for the following two reasons.
The first one is that
the present  mass of the  SMC ($\sim 3 \times 10^9 {\rm M}_{\odot}$)
is significantly smaller than that of the LMC 
($2 \times 10^{10} {\rm M}_{\odot}$;  GN96):
this means that the present mass of the SMC is much smaller
than the original mass of the LMC ($\sim 10^{11} {\rm M}_{\odot}$).
The second is that  we do not intend
to discuss the formation of the Magellanic Stream from the tidal stripping
of the gas initially in the SMC.  We consider that the inclusion of the SMC
can hardly change the present results on the influences of the LMC on
the dynamical evolution and the star formation history of the Galaxy. 
Brief discussion of the orbital evolution of the LMC and the SMC
in a live  Galactic potential is given in Bekki (2011).

\begin{table}
\begin{minipage}{80mm}
\caption{The particle mass and the total
particle number of each component adopted for
the standard model.}
\begin{center}
\begin{tabular}{ccc}
{Component} & {Mass} & {Number} \\
{} & {} & {} \\
{The Galaxy} & {} & {} \\
{Dark matter} & { $1.3 \times 10^6 {\rm M}_{\odot}$ } & 800000 \\
{Stellar disk} & { $2.9 \times 10^5 {\rm M}_{\odot}$ } & 200000 \\
{Bulge} & { $2.9 \times 10^5 {\rm M}_{\odot}$ } & 33400 \\
{Extended disk} & { $1.2 \times 10^5 {\rm M}_{\odot}$ } & 50000 \\
{} & {} & {} \\
{LMC} & {} & {} \\
{Dark matter} & { $9.0 \times 10^5 {\rm M}_{\odot}$ } & 100000 \\
{Stellar disk} & { $5.4 \times 10^4 {\rm M}_{\odot}$ } & 100000 \\
\end{tabular}
\end{center}
\end{minipage}
\end{table}

\begin{figure}
\psfig{file=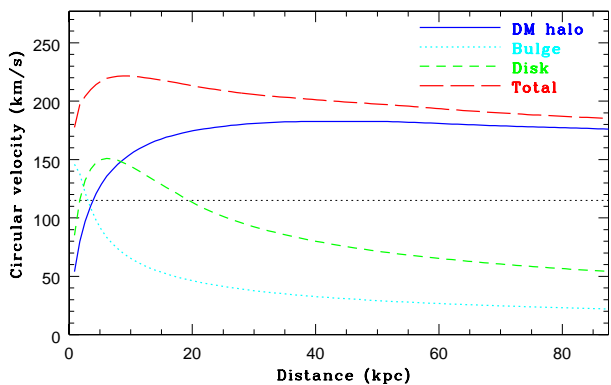,width=8.0cm}
\caption{
The rotation curve profile as a function of radius
in the Galaxy  for  the standard model (red long-dashed line). Contributions
of the dark halo, stellar bulge, and stellar disk are shown by
blue solid line, cyan dotted one, and  green short-dashed one, respectively.
The black dotted line represents the maximum circular velocity of the
LMC  with $M_{\rm dm, l}=9 \times 10^{10} {\rm M}_{\odot}$
in the standard model.
}
\label{Figure. 1}
\end{figure}

\begin{figure}
\psfig{file=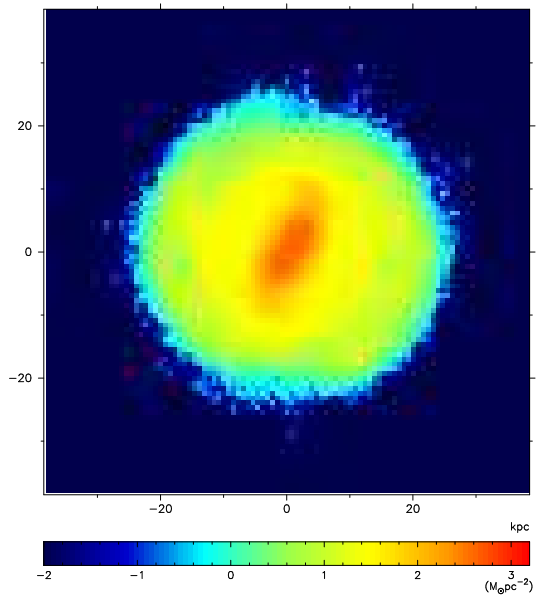,width=7.0cm}
\caption{
The 2D distribution of $\mu_{\rm s}$ projected onto the $x$-$y$ plane
for disk stars of the Galaxy in the isolated model. Here
$\mu_{\rm s}$ is the smoothed surface mass density of stars in logarithmic
scale (i.e., $\mu_{\rm s} =\log_{10} \Sigma_{\rm s}$, where
$\Sigma_{\rm s}$ is the surface mass density of the stars).
The 2D distribution is constructed based on the simulation data sets
at $T=5.2$ Gyr and the Gaussian smoothing of 875pc. 
}
\label{Figure. 2}
\end{figure}

\begin{figure*}
\psfig{file=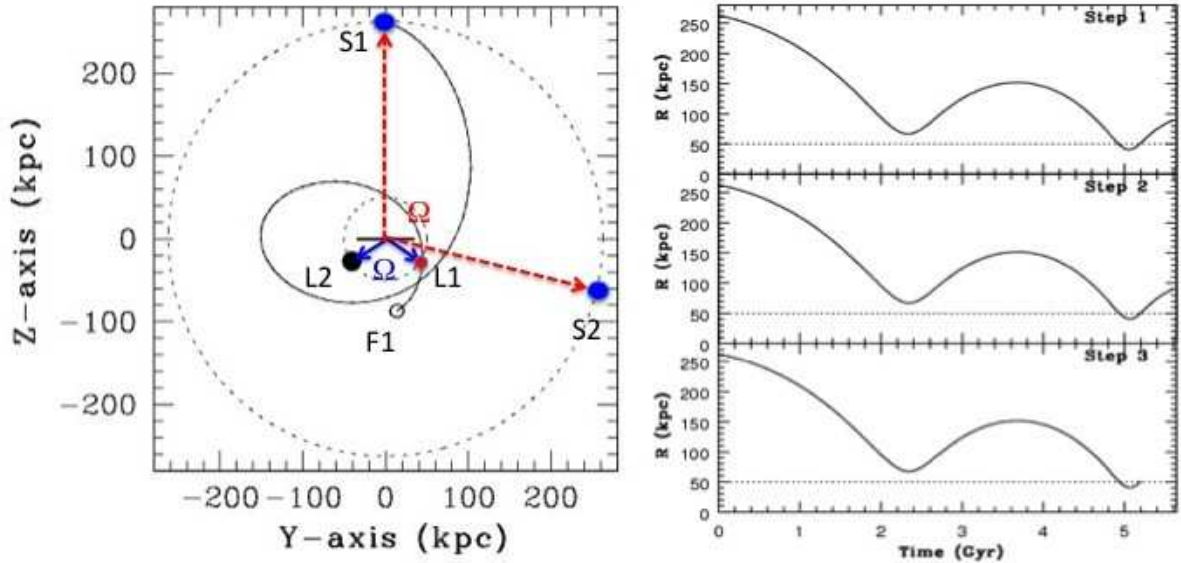,width=14.0cm}
\caption{
An illustrative figure for the new coordinate transformation method
(CTM) by which we can reproduce the present location of the LMC rather
precisely for a given set of orbital parameters of the LMC.
In the left panel, the initial and final location of the LMC
in the high-resolution simulation at Step 1 of the CTM are shown
by a filled blue circle (S1) and a open circle (F1), respectively.
The observed location of the LMC is shown by a filled black circle
(L2) and the location L1 describes the point where the simulated
LMC reaches $R=50$ kpc for the second time in the simulation 
at Step 1. The angular distance between L1 and L2 is denoted
as $\Omega$ and should be the same as that between S1 and S2, where
S2 is the initial location of the LMC in the simulations at
Step 2 and Step 3.
The virial radius of the Galaxy and the present distance of the LMC
from the Galaxy are shown by  large and small dotted lines,
respectively, in the left panel.
In the right panel, the orbital evolution of the LMC is shown for the
three simulations at  Step 1 (top), Step 2 (middle), and
Step 3 (bottom). The present distance of the LMC is shown by a dotted
line in this panel. The details of the CTM are given in the main text.
}
\label{Figure. 3}
\end{figure*}

\subsection{Simulation setup}

Numerical computations
were carried out both  on
(i) the latest version of GRAPE
(GRavity PipE, GRAPE-DR) -- which is the special-purpose
computer for gravitational dynamics (Sugimoto et al. 1990)
and (ii)  five high-end PCs and servers  (e.g., one IBM system iDataPlex)
with one or two GPU cards (Tesla M2050 and GTX580) 
and CUDA G5/G6 software package being
installed for calculations of gravitational dynamics
at University of Western Australia. 
We adopt a direct summation method for the calculation
of gravitational force of each individual particle.
It takes about 23 CPU hours
for a GRAPE-DR machine to finish
a high-resolution simulation for 8.4 Gyr evolution
(corresponding to 3000 time steps with $2.8 \times 10^6$ yr
time interval)  of the LMC-Galaxy
system.  Although the calculation speed for a simulation carried out
by a GPU machine is significantly slower than that by the GRAPE,
the five GPU machines enable us to investigate a parameter space for
the LMC orbit in a quite efficient and productive  way.

\subsection{A new method to find an orbital  model that reproduces the
present LMC position}

The present study adopts the new ``coordinate transformation
method (CTM)'' rather than the backward integration scheme adopted
by many previous studies (e.g., MF80; GN96; BC05; B07; DB11a, b, c)
so that we can (i) investigate both the mass loss of the LMC
and the dynamical friction of the LMC self-consistently
and (ii) reproduce the present location of the LMC rather precisely.
The present positions of the sun and the LMC are  ($-8.5$, 0, 0) kpc
and (-1.0, -40.8, -26.8) kpc, respectively (e.g., GN96; BC05).
The sun is currently moving to the direction of the positive $y$ 
in the coordinate system of the present study. This means that
the Galaxy is rotating clockwise if it is viewed from the north Galactic
pole (GN96).
The initial distance between the LMC and the Galaxy is a free parameter
and described as $R_{\rm i}$. For all models,  $R_{\rm i}=15R_{\rm d}$
($>R_{\rm vir}$) so that we can investigate the orbital evolution of the LMC
from outside the virial radius of the Galaxy. 
The tangential and radial velocities of the LMC at $R=R_{\rm i}$ are
set to be $f_{\rm v,t}v_{\rm cir}$ and $f_{\rm v,r}v_{\rm cir}$,
respectively, where $v_{\rm cir}$ is the circular velocity at $R_{\rm i}$:
The LMC has a velocity of $f_{\rm v} v_{\rm cir}$ at $R_{\rm i}$,
where $f_{\rm v}$=$(f_{\rm v,t}^2+f_{\rm v,r}^2)^{0.5}$.

In order to more clearly explain the CTM in Fig.3, we here consider that
the orbital plane of the LMC is coincident with the $y$-$z$ plane
and the LMC has the initial velocity of 
(0,$f_{\rm v,t}v_{\rm cir}$, $-f_{\rm v, r}v_{\rm cir}$) at $R_{\rm i}$.
Fig. 3 illustrates the CTM for the LMC's orbital plane being the same
as the $y$-$z$ plane.
We take the following four steps 
(Step 0, 1, 2, and 3) in the CTM to find an orbital model
that can reproduce the present LMC position for a given set of
model parameters.

\subsubsection{Step0: Running low-resolution simulations}

First, we run a number of low-resolution simulations for a given
set of model  parameters 
(e.g. , $M_{\rm dm, l}$,
$R_{\rm i}$, $f_{\rm v,t}$, and $f_{\rm v, r}$) in order to
find  how long it takes for the LMC to have $R=50$ kpc 
for the {\it second} time.  Here the present LMC is considered to be going
away from the Galaxy after the last  pericenter passage
($R<50$ kpc)  about 0.2 Gyr ago.
The time when the simulation starts is set to be 0 whereas
the time when the LMC passes the pericenter for the second time is 
referred to as $T_{\rm 50}$ for convenience. 
The particle numbers for dark matter, disk, bulge, and EDC  of 
the Galaxy, and dark matter and disk of the LMC 
in a low-resolution simulation
are 200000, 50000, 8350, 20000, 10000, and 10000, respectively.
Although the total particle number of the low-resolution
simulation is   $\sim 300000$, this number is enough
to estimate $T_{\rm 50}$. For models
with smaller $M_{\rm dm, l}$ and larger $f_{\rm v}$,
$T_{\rm 50}$ can be quite long ($>8$ Gyr).

\subsubsection{Step1: Running high-resolution simulations}

If $T_{\rm 50}$ is determined, we run a  high-resolution model
with $N>1000000$ for a time scale ($T_{\rm f}$)
significantly larger than  $T_{\rm 50}$ derived in the low-resolution
one.
This is mainly because even if the initial location 
of the LMC (``S1'') is 
the same between the low- and high-resolution simulations,
the final location (``F1'') can be different between the two simulations.
Thus we run a high-resolution simulation for $T_{\rm f}$
to ensure that the LMC can pass $R=50$kpc  at least twice.
The orbital evolution of the LMC in this high-resolution model
is recorded for the following steps.

\begin{table}
\centering
\begin{minipage}{80mm}
\caption{$T_{50}$ and $\Omega$ for
the representative four models estimated by using the new CTM.}
\begin{center}
\begin{tabular}{ccc}
{Model}
& {$T_{50}$
\footnote{The time ($T$) when the LMC passes through
the pericenter ($R<50$ kpc) for the second time. 
The definition of this quantity is given 
in the main text.}}
& {$\Omega$
\footnote{The angular distance between the original
and the transformed locations for the LMC position (at $T=0$ in
Figure 3).
The definition of this quantity is given 
in the main text and in Figure 3.}} \\
T1 & 5.2 Gyr & 334$^{\circ}$  \\
T2 & 2.2 Gyr & 241$^{\circ}$   \\
T3 & 2.3 Gyr & 271$^{\circ}$   \\
T7 & 7.4 Gyr & 5$^{\circ}$   \\
\end{tabular}
\end{center}
\end{minipage}
\end{table}

\begin{figure}
\psfig{file=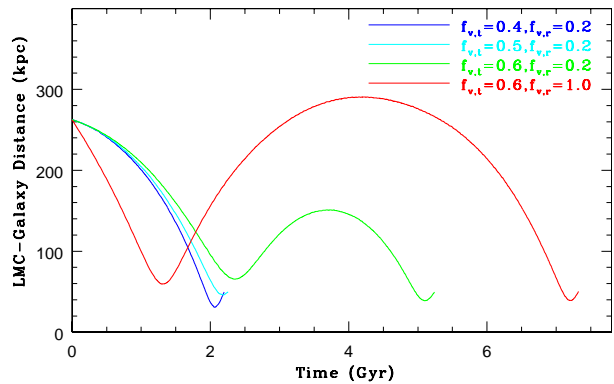,width=8.0cm}
\caption{
The orbital evolution of the LMC in
the four representative models that can reproduce the present
location of the LMC: 
$f_{\rm v,t}=0.4$ and $f_{\rm v,r}=0.2$ (blue),
$f_{\rm v,t}=0.5$ and $f_{\rm v,r}=0.2$ (cyan),
$f_{\rm v,t}=0.6$ and $f_{\rm v,r}=0.2$ (green), and
$f_{\rm v,t}=0.6$ and $f_{\rm v,r}=1.0$ (red).
}
\label{Figure. 4}
\end{figure}

\begin{figure}
\psfig{file=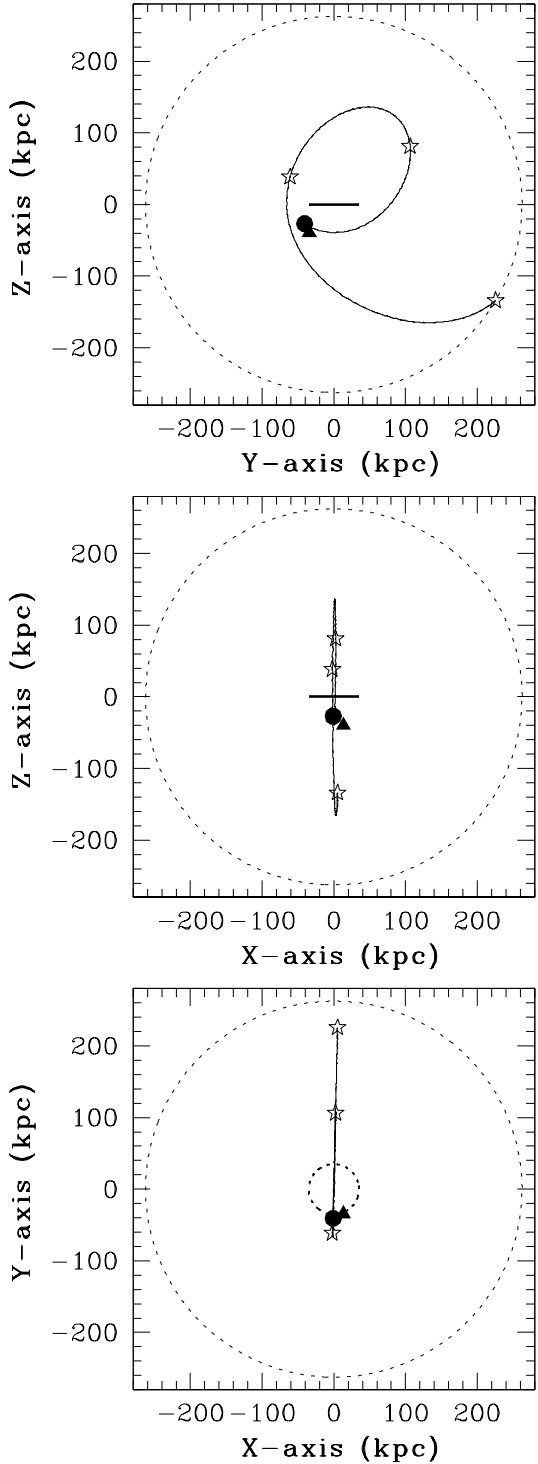,width=7.0cm}
\caption{
The orbital evolution of the LMC projected onto the $y$-$z$ plane
(top), the $x$-$z$ one (middle), and the $x$-$y$ one (bottom) 
in the standard model. The size of the EDC and the virial radius
of the Galaxy are shown by a  thick solid line and a thin dotted
one, respectively. The size of the EDC is shown
by a dotted line in the $x$-$y$ projection so that
the locations of the LMC and the SMC can be more clearly seen.
The present locations of the LMC and the SMC
are shown by a circle and a triangle, respectively.
The locations of the simulated LMC at $T=-5.2$ Gyr, $-2.7$ Gyr,
and $-1.0$ Gyr are shown by open stars.
}
\label{Figure. 5}
\end{figure}

\begin{figure}
\psfig{file=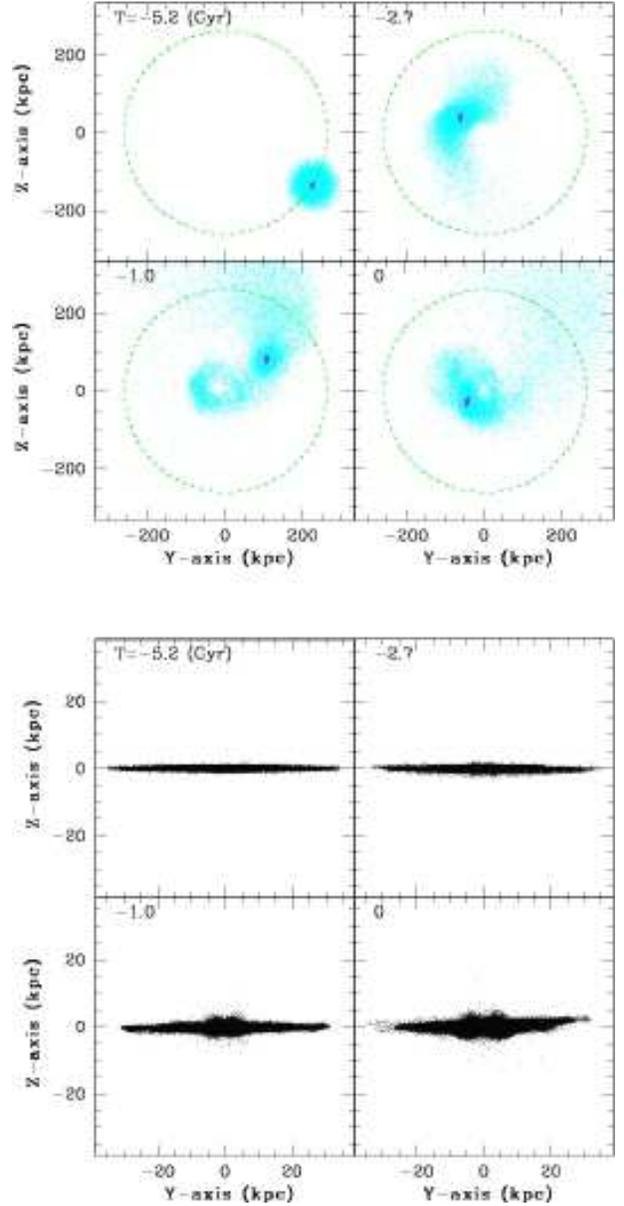,width=8.0cm}
\caption{
The evolution of the mass distribution
projected onto the $y$-$z$ plane for
the LMC (upper four) and the EDC of the Galaxy (lower four)
in the standard model. The disk and dark halo components of the LMC
are shown by red and cyan particles, respectively,  in the upper panel,
and the dotted line represents the virial radius of the Galaxy.
The time in the upper left corner of each panel is given in units of
Gyr.
}
\label{Figure. 6}
\end{figure}

\begin{figure}
\psfig{file=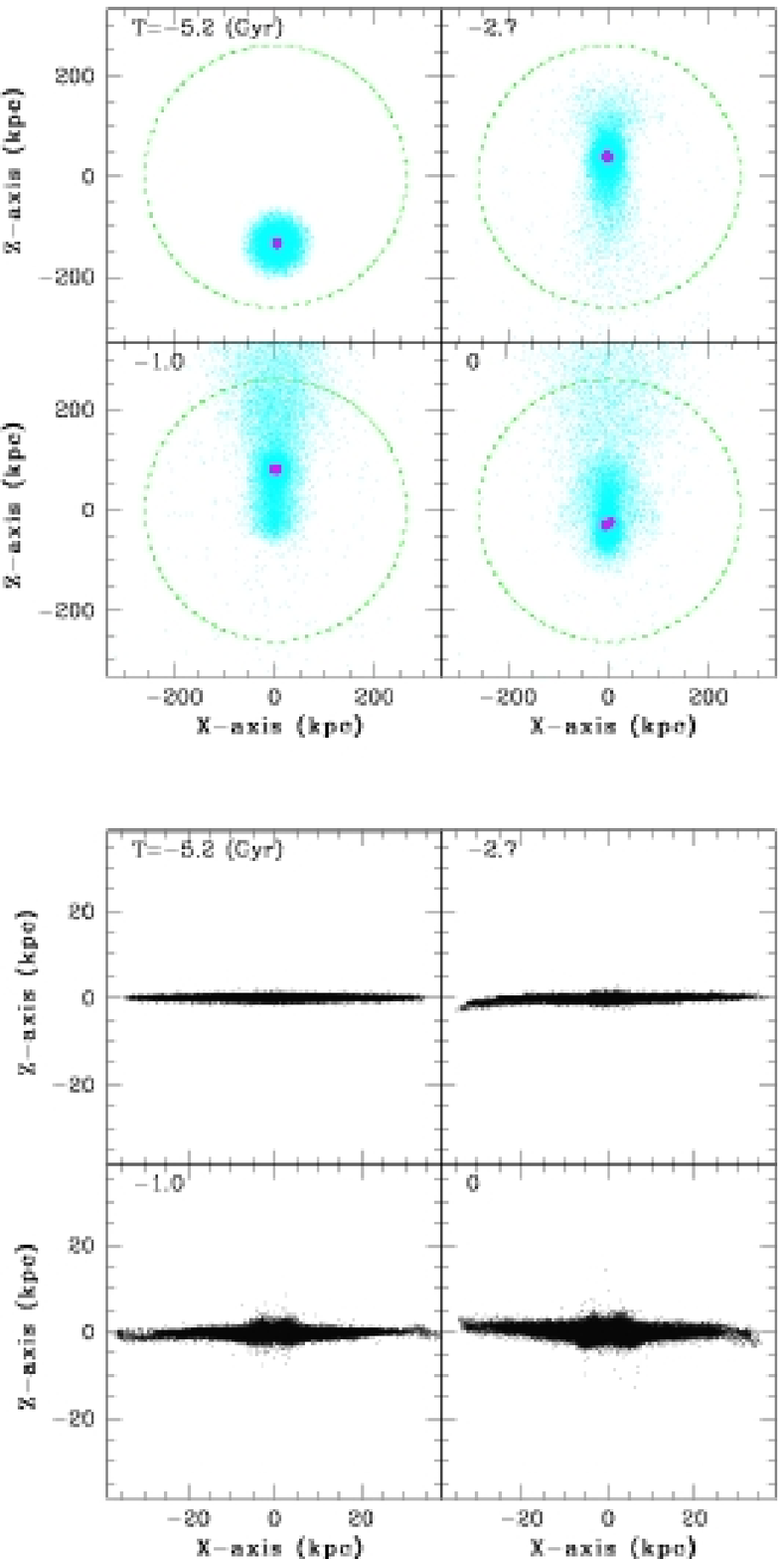,width=8.0cm}
\caption{
The same as Fig. 6 but for the $x$-$z$ projection.
}
\label{Figure. 7}
\end{figure}

\subsubsection{Step 2: Coordinate transformation}

In this Step 2, we investigate the 3D position 
of the LMC (``L1'' point in Fig. 3) in
the Galactic coordinate when the LMC  passes  $R=50$ kpc for the 
second time ($T_{50}$)  in the high-resolution
simulation.  We compare the differences in the 3D
position of the LMC between the simulation (``L1'') 
and the observation (``L2'') and thereby estimate the angular distance
($\Omega$) between  L1 and L2.
We rotate the initial
location of the LMC (``S1'') in the simulation  by $\Omega$
so that the initial location of the LMC can be ``S2''.
We again run a  high-resolution simulation with  a new initial location
of the LMC (S2) for $T_{\rm f}$ 
and thereby try to find  when the LMC 
passes $R=50$ kpc for the second time ($T_{\rm 50}$). 

\subsubsection{Step4: Rerunning simulations}

If we can confirm whether the simulated LMC's position at $T_{50}$  
can be almost identical to the observed location of the LMC,
we rerun a high-resolution simulation (for the initial
position S2) only for $T_{\rm 50}$
and record the simulation data at the final time step ($T=T_{50}$)
in order to analyze the data.  The difference
between the simulated and the observed locations 
($\Delta r=(\Delta x^2+ \Delta y^2 + \Delta z^2)^{0.5}$) 
can be  well less than $\sim 2$ kpc
in the CTM. 
Thus, we have to run at least 4 simulations to have a model that
can reproduce the present location of the LMC rather precisely
for a given set of model parameters.

\subsubsection{Advantages and disadvantages}

The underlying assumption in the CTM is that the
outer part of the Galaxy can be regarded as
having an almost spherical mass distribution.
Owing to the inclusion of a stellar disk in the present model, the
Galactic potential is triaxial to some extent
in the inner region of the Galaxy. 
However the CTM enables us to derive the LMC orbit model that can
reproduce the present LMC position quite well,  because both the apocenter
and the pericenter of the LMC are significantly larger than
the disk size of the Galaxy. 
Thanks to the CTM, we do not have to run a huge number of
high-resolution simulations to find a model that reproduces the present
LMC position quite well. This new CTM can be used for investigating
the past orbital evolution of  other dwarf galaxies in the Local Group.
One of disadvantages of the CTM is that it can be difficult to
reproduce the present location of the LMC for the 
case where the Galactic potential is triaxial to a significant extent.
This weakness of the CTM needs to be improved in our future studies.

Also, although $\Delta r$ can be rather small in the CTM,
the velocity difference between the observed and simulated one
($\Delta v=(\Delta v_{\rm x}^2+ \Delta v_{\rm y}^2 + 
\Delta v_{\rm z}^2)^{0.5}$) 
can not be so small.  For example,  $\Delta v$ in the standard model (T1) 
is $\sim 80$ km s$^{-1}$ (i.e., $\sim 40$ km s$^{-1}$ in each velocity
component). Other models show similar magnitudes of 
velocity differences ($\Delta v$). The difference 
in the proper motion of the LMC
between different observations (e.g., Kallivayalil et a. 2006; 
Costa et al. 2009; Vieira et al. 2010), 
is not so small ($\sim 80$ km s$^{-1}$) either.
If future observations will more precisely determine the 
proper motion of the LMC (with an accuracy of less than 10 km s$^{-1}$),
then the CTM needs to be improved significantly.

\subsubsection{The orbital plane of the LMC}

The orbital plane of the LMC is defined as a plane that is perpendicular
both to the position vector of the LMC and the velocity one of the LMC.
In the present study, we consider that the orbital plane of the LMC is
similar to those used in successful models for the formation of the
Magellanic Stream (e.g., GN96, DB11c): the best model by GN96 has 
the present LMC velocity of ($-5$, $-225$, $194$) km s$^{-1}$.
Fig. 4 shows four examples of successful models
(T1, T2, T3, and T7) 
in which the orbital planes are almost the  same as that in GN96
and the present location of the LMC can be well reproduced.
Table 3 summarizes $T_{50}$ and $\Omega$ in the CTM for these four models.
We also investigate some models in which the orbital plane of the LMC
is the same as that 
adopted in DB11c, in which the present 3D velocity
of the LMC is ($-51$, $-226$, 229) km s$^{-1}$.
We describe the results of the models with the best orbit in GN96,
because the results of the models with the two different orbital
planes adopted by GN96 and DB11c are essentially similar.

\subsection{Estimation of  the Galactic precession/nutation/pole-shift}

We investigate the time evolution of the rotation (i.e. angular momentum)
vector ${\bf L}_{\rm d}$ =
($L_{\rm x}$, $L_{\rm y}$, $L_{\rm z}$) of the stellar disk for each model:
this vector defines the orientation of the  rotation axis of the stellar disk.
We investigate
(i) the  inclination angle ${\theta}_{\rm d}$ between
the $z$-axis and the rotation vector of the disk galaxy
and (ii) the angle ${\phi}_{\rm d}$ between
the rotation vector projected onto the $x$-$y$ plane
and the $x$-axis
at selected time steps
for each model.
We estimate ${\theta}_{\rm d}$ and ${\phi}_{\rm d}$  as $\arccos(L_{\rm z}/L_{\rm t})$
and $\arccos(L_{\rm x}/L_{\rm xy})$, respectively,
where $L_{\rm t}$ is the absolute magnitude of the angular momentum
and $L_{\rm xy}={({L_{\rm x}}^2+{L_{\rm y}}^2)}^{0.5}$.
$L_{\rm z}$ is positive if the Galaxy rotates clock-wise viewed from
the north Galactic pole.
The rotation vector ${\bf L}_{\rm d}$ is estimated with respect to
the central stellar particle that is initially located at the exact center of
the disk and  has no initial 3D velocities.
However, these central particles
later can be slightly dislocated from their initial central
positions and  have tiny 3D velocities due to
the evolution of the disk. As a result of this, the time evolution
of {\bf L}$_{\rm d}$ derived in some models  can show rapid short-term
increase
and decrease with a small amplitude.

We also investigate the rate of precession/nutation/pole-shift of the Galaxy
($\dot{ {\theta}_{\rm d} }$) in order
to provide predictions of the present $\dot{ {\theta}_{\rm d} }$ of the 
Galaxy. The predicted $\dot{ {\theta}_{\rm d} }$ can be compared with
the corresponding observations, if future observational studies based
on GAIA and VLBI can detect the Galactic precession/nutation/pole-shift. 
Although $\dot{ {\theta}_{\rm d} }$ can become both negative and positive
in its long-term evolution,
$\dot{ {\theta}_{\rm d} }$ can keep its sign
and does not change significantly  for the last $\sim 50$ Myr
for most models.
This implies that a more robust comparison between the predicted
present $\dot{ {\theta}_{\rm d} }$ and the observed one can be done.

The angular momentum vector ({\bf L}$_{\rm d}$)
is derived by using all disk particles within 21 kpc of the Galaxy
in the present study.
Accordingly it is possible that ${\theta}_{\rm d}$  can be due largely  to
the outer warp formed in the disk during the LMC-Galaxy interaction.
We investigate  {\bf L}$_{\rm d}$ for $R=3.5$ kpc, 8.5 kpc, and 21 kpc 
of the Galaxy
at the final time step
in the standard model 
and confirm that the angle between  
{\bf L}$_{\rm d}$ for $R=8.5$ (3.5) kpc and 21 kpc is only
$0.02^{\circ}$ ($0.001^{\circ}$). This means that 
if the Galaxy shows the pole shift, then the entire disk shows
the pole shift (i.e., not simply due to the warp).

\subsection{Parameter study}

Although we have investigated a very 
large number of models, we describe the results
of the 27 representative models in the present study. The orbital
plane of the LMC is almost the same as that adopted in previous works
by GN96 for all  models.
The Table 1 summarizes the model parameters
for these representative models (T1$\sim$24) and for isolated models
(I1$\sim$3). No LMC-Galaxy interaction is included in the three isolated
models. 
Models in which the LMC experiences only 
one pericenter passage are referred to as "first passage" models, 
while those in which it experiences more are referred to as
"multiple-passage" models.

We mainly describe the results of the ``standard model'' (T1), because
we think that (i) the model parameters are quite reasonable and 
(ii) it shows
more clearly the possible influences of the LMC on the evolution
of the Galaxy. 
The LMC can not show $R=50$ kpc within $\sim 8$ Gyr in some models
with lower total masses of the LMC
so that the result of these models are not useful in discussing the
influences of the LMC on the Galaxy. However these models are quite
useful in discussing the possible minimum LMC mass above which 
the LMC can have $R=50$ kpc within a reasonable time scale.
We thus discuss  the results of these models in the Appendix A.

The EDC is much more significantly influenced by the LMC 
in comparison with the main stellar disk. We therefore
mainly discuss the final structure and kinematics of the EDC
in each model. 
We consider that the Galactic precession/nutation/pole-shift
by distant tidal encounters
between the Galaxy and its satellite galaxies are quite important in
the Galactic astrometry.
Therefore, we discuss the results of the models in which the  parameter values
are not reasonable for the LMC but possible for other satellite galaxies
and previous infall of small groups
in the Appendix B. Since we have already described the 3D distribution
of stars stripped from the LMC stellar halo and discussed the importance
of the distribution in understanding the past 3D orbit of the LMC
in Bekki (2011),
we only briefly discuss the distribution in the Appendix C.

\begin{figure}
\psfig{file=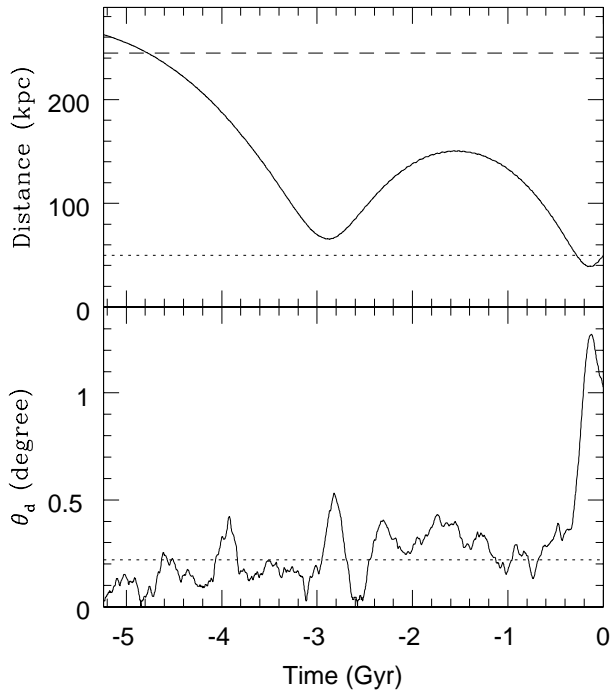,width=8.0cm}
\caption{
The time evolution of the distance between the LMC and the Galaxy
(upper) and $\theta_{\rm d}$ (lower) for the standard model (T1).
The present LMC-Galaxy distance and the virial radius of the Galaxy
are shown by dotted and dashed lines, respectively, in the upper
panel. The maximum value of $\theta_{\rm d}$ for $\sim 2$ Gyr isolated
evolution of the Galaxy in the model I1
is shown by a dotted line in the lower panel
for comparison.
}
\label{Figure. 8}
\end{figure}

\begin{figure}
\psfig{file=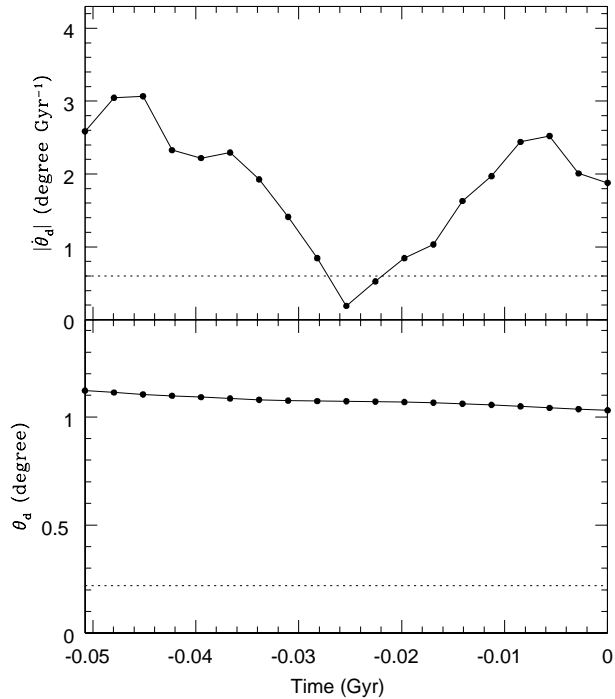,width=8.0cm}
\caption{
The time evolution of 
$\dot{ {\theta}_{\rm d} }$ (upper) and $|\theta_{\rm d}|$ (lower)
for the last $\sim 0.05$ Gyr in the standard model.
The maximum values of 
$\dot{ {\theta}_{\rm d} }$  and
$\theta_{\rm d}$ 
for $\sim 2$ Gyr isolated
evolution of the Galaxy in the model I1
are  shown by dotted lines
for comparison.
}
\label{Figure. 9}
\end{figure}

\begin{figure}
\psfig{file=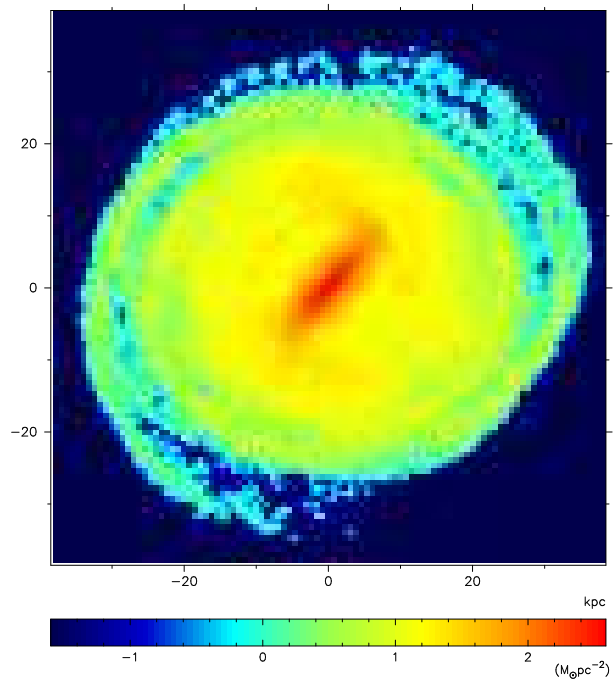,width=8.0cm}
\caption{
The same as Fig. 2 but for the present  EDC
(i.e., 0 Gyr) of the Galaxy in
the standard model.
}
\label{Figure. 10}
\end{figure}

\begin{figure}
\psfig{file=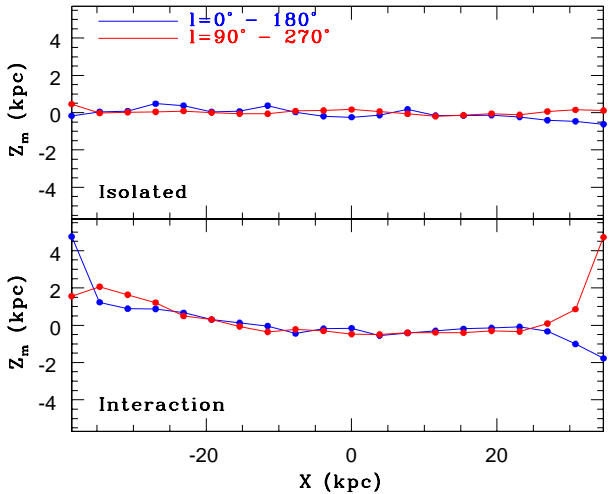,width=8.0cm}
\caption{
The mean $z$-position ($Z_{\rm m}$) for the stars in  the EDC 
of  the Galaxy as a function of the projected distances ($X$)
for the isolated model I1 (upper) and the standard interaction 
model T1 (lower).
The blue and red lines show the $Z_{\rm m}$ profiles for
the $x$-$z$ projection (i.e., for $l=0^{\circ} - 180^{\circ}$)
and the $y$-$z$ one (i.e., for $l=90^{\circ} - 270^{\circ}$),
respectively.
}
\label{Figure. 11}
\end{figure}

\section{Results}
\subsection{The standard model}

Fig. 5 shows the orbital evolution of the LMC with respect to the Galactic
center in the standard LMC-Galaxy interaction model T1. 
The radial scale length and vertical scale height of the EDC in this model
are $0.4R_{\rm d}$ (7 kpc) and $0.02R_{\rm d}$ (0.35 kpc), respectively.
The LMC passes its pericenter ($R=65$ kpc) for the first time
about 2.9 Gyr ago and for the second time about 0.2 Gyr ago in this model.
The orbit of the LMC is almost an polar one and  
its  orbital plane of the LMC is almost identical to the $y$-$z$ plane. 
Fig. 6 shows that the distributions of the  dark matter halo of the LMC
at four representative time steps. After the first pericenter passage,
the LMC can lose a significant fraction of its dark matter halo to
the outer part of the Galaxy. About $\sim 33$\% of the halo 
within $10R_{\rm d, l}$ cab be  stripped
by the strong tidal field of the Galaxy within the latest 5.2 Gyr.
The stripped dark halo can form a ring-like distribution in the Galactic
halo and could possibly influence the orbital evolution of other Galactic
satellite galaxies.

Figs. 6 and 7 show how the first and the second LMC-Galaxy tidal encounters
(i.e., the two pericenter passages)
dynamically influences the EDC of the Galaxy. 
The first LMC-Galaxy encounter can only slightly influence the very outer
part ($R>30$ kpc) of the EDC: the outer EDC is slightly warped in the
$x$-$z$ plane. After the second LMC-Galaxy encounter,
the present  EDC of the Galaxy projected onto the $x$-$z$ and
$y$-$z$ planes
shows a clear sign of warps in its outer part ($R>20$ kpc).
The EDC projected onto the $y$-$z$ plane appears
to have a   ``bow-like'' morphology (i.e., the two edges of the EDC are 
in the regions with $z>0$). Given that the isolated disk model (I1) does
not show such warp-like structures (later shown in Fig. 11), 
these results clearly demonstrates
that the LMC-Galaxy tidal interaction  can induce warps in the outer
part of the EDC.  This is confirmation of the results of previous
simulations that show the formation of the Galactic warps by
the LMC-Galaxy interaction (e.g, Tsuchiya 2002;
Weinberg \& Blitz 2006).

As shown in Figs. 6 and 7,  the disk stars of the LMC can not be stripped
by the strong tidal field of the Galaxy, though the stellar disk
can be dynamically heated up to form a thick disk.
This no stripping of the LMC disk stars is due to the fact that
the original LMC mass within $2R_{\rm d, l}$ (=15 kpc) is 
$\sim 4 \times 10^{10} {\rm M}_{\odot}$ for the LMC to 
 have the maximum circular
velocity of $\sim 110$ km  s$^{-1}$.
If the original mass of the LMC is significantly smaller,
then the stellar disk of the LMC can be stripped, as shown in the low-mass
LMC model by BC05.
The LMC can develop a strong stellar bar
in the inner region of the disk during the LMC-Galaxy tidal interaction.
The inner part of the LMC ($<2R_{\rm d, l}=35$ kpc) can keep about 85\%
of its original mass after the 5.2 Gyr LMC-Galaxy tidal interaction.

Fig. 8 shows the time evolution of the LMC-Galaxy distance
and $\theta_{\rm d}$.
Clearly the inclination angle of the stellar disk
in the Galaxy  can suddenly  increase and decrease
before and after the first pericenter passage about $\sim 3$ Gyr ago.
Although the change of $\theta_{\rm d}$ is small (i.e., less than one degree),
this sudden change may well be better described as
``pole shift'' rather than regular precession/nutation.
 After the first pericenter 
passage, the stellar disk of the Galaxy can continue to 
keep a higher  $\theta_{\rm d}$ till $0.4$ Gyr ago. 
The disk can increase significantly and suddenly
 $\theta_{\rm d}$ during 
second pericenter passage owing to the gravitational torque
of the LMC.  The disk decrease  $\theta_{\rm d}$ after
the second pericenter passage and finally  it has 
$\theta_{\rm d}=1.0^{\circ}$ and $\phi_{\rm d}=125.5^{\circ}$ (at the present).

Fig. 9 shows the time evolution of $\theta_{\rm d}$ and the change rate
($\dot{ {\theta}_{\rm d} }$) for the last $\sim 0.05$ Gyr. The value of
$\dot{ {\theta}_{\rm d} }$ at the final time step can correspond to
the present pole-shift  rate of the Galaxy. Here we use the term
``pole-shift rate'' just for convenience and clarity, though the 
long-term precession/nutation  is not a regular one. 
Clearly, $\theta_{\rm d}$ decreases quite steadily
for the last 0.05 Gyr, though the time evolution of $\dot{ {\theta}_{\rm d} }$
is slightly irregular. The negative values of $\dot{ {\theta}_{\rm d} }$
for the last 0.05 Gyr 
strongly suggest that the Galactic disk will continue to decrease
$\theta_{\rm d}$ in the future (at least till the LMC passes
its third pericenter).
The present pole-shift rate is estimated as
1.9 degree Gyr$^{-1}$ corresponding to $\sim 6.8 \mu$as yr$^{-1}$
in this standard model.
The derived  $\dot{ {\theta}_{\rm d} }$ in the standard model
is significantly larger than 
that (at most 0.6  degree Gyr$^{-1}$) 
derived for the isolated model I1. Therefore the higher pole-shift 
rate can be regarded as being due to the LMC-Galaxy interaction.

Fig. 10 shows that the distribution of the EDC projected onto the $x$-$y$ plane
(i.e., the Galactic plane) appears to be significantly elongated 
and significantly distorted. 
Furthermore, the outer part of the EDC has two distinct spiral-arm
structures, which were formed as a result of the last LMC-Galaxy tidal
interaction. These spiral-arm structures might well correspond to
the distant (18 kpc $<R<$ 24 kpc)
extended spiral arms  of the Galaxy discovered by 
McClure-Griffiths et al. (2004).
The central bar in the EDC is formed not from the LMC-Galaxy
interaction but from the dynamical interaction between the newly developed
stellar bar of the Galaxy and the EDC. These results strongly suggest that
the outer part ($R>20$ kpc) of 
the Galactic gas disk can have a significantly asymmetric
distribution owing to the LMC-Galaxy tidal interaction.  
Thus both the vertical structure of the Galaxy (e.g., the presence of
the warps) and the non-asymmetric structure of the outer Galaxy have
fossil information of the past LMC-Galaxy interaction history.

Fig. 11 shows the mean $z$-position ($Z_{\rm m}$) of the stars in the EDC
as a function of the projected positions ($X$). 
In the Galactic coordinate system adopted in the present study,
the location of the Sun is set to be ($-8.5$, 0. 0) kpc.
Therefore the positive (negative) direction of the $x$-axis points to  
$l=0^{\circ}$ ($l=180^{\circ}$) whereas the positive (negative)
direction of the $y$-axis points to 
$l=90^{\circ}$ ($l=270^{\circ}$). 
The $Z_{\rm m}$ profile for
the stars in the EDC  projected onto the $x$-axis 
clearly shows a warp with the morphology similar to an ``integral symbol'':
the warp structure has a positive $Z_{\rm m}$ for $x<-20$ kpc
and a negative $Z_{\rm m}$ for $x>20$ kpc.  
The $Z_{\rm m}$ profile 
for the stars projected onto the $y$-axis 
also shows a warp with a ``bow-like'' shape  (i.e.,
positive $Z_{\rm m}$ both for $y<-20$ kpc and for $y>20$ kpc).
These results demonstrate that the shape of the Galactic bar can depend
on from where it is observed.

The LMC-Galaxy interaction can also influence the kinematics of the
EDC in the Galaxy. The present vertical velocity dispersion 
($\sigma_{\rm z}$) of the EDC
at $R=26$ kpc in the isolated model is 4.0 km s$^{-1}$ whereas
it is 5.7 km s$^{-1}$ in the standard model 
(i.e., with LMC-Galaxy interaction).
The difference in $\sigma_{\rm z}$ between the two models
can be also clearly seen for $R>20$ kpc. For example,
$\sigma_{\rm z}$ at $R=35$ kpc in the isolated model is 2.8 km s$^{-1}$ whereas
it is 3.4 km s$^{-1}$ in the standard model. 
These results clearly demonstrate that the outer kinematics
of the extended stellar and gaseous disks of the Galaxy have fossil
information on the past LMC-Galaxy tidal interaction.


\begin{figure}
\psfig{file=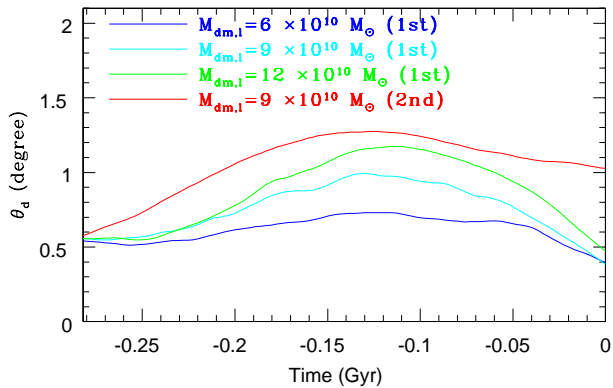,width=8.0cm}
\caption{
The time evolution of $\theta_{\rm d}$ for the last 0.28 Gyr in
the four different models: 
T19 with $M_{\rm dm,l}=6 \times 10^{10} {\rm M}_{\odot}$ 
and $f_{\rm v}=1.12$ (blue),
T18 with
$M_{\rm dm,l}=9 \times 10^{10} {\rm M}_{\odot}$ and $f_{\rm v}=1.12$ (cyan),
T20 with
$M_{\rm dm,l}=12 \times 10^{10} {\rm M}_{\odot}$ and $f_{\rm v}=1.12$ (green),
and T1
$M_{\rm dm,l}=9 \times 10^{10} {\rm M}_{\odot}$ and $f_{\rm v}=0.63$ (red).
The first three models are first passage models with
higher present velocities of the LMC. The model with $f_{\rm v}=0.63$ 
is the standard model.
}
\label{Figure. 12}
\end{figure}

\begin{figure}
\psfig{file=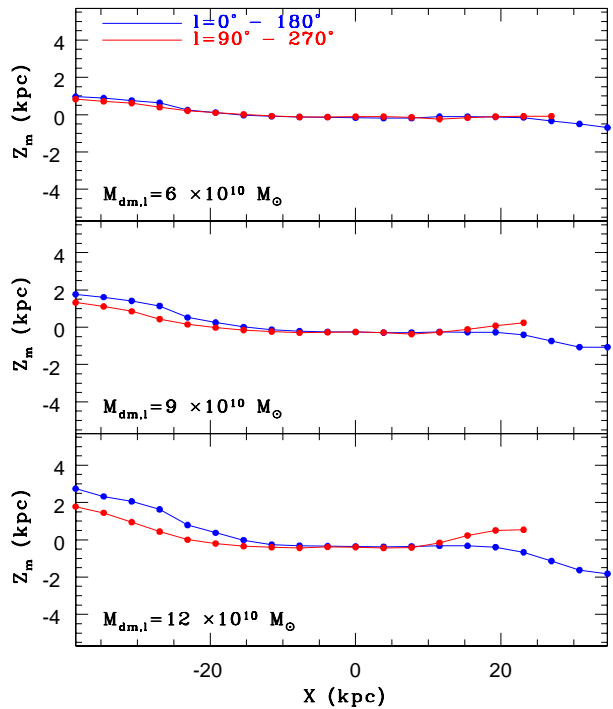,width=8.0cm}
\caption{
The same as Fig. 11 but for three different first passage models
with 
$M_{\rm dm,l}=6 \times 10^{10} {\rm M}_{\odot}$ (top),
$M_{\rm dm,l}=9 \times 10^{10} {\rm M}_{\odot}$ (middle), and
$M_{\rm dm,l}=12 \times 10^{10} {\rm M}_{\odot}$ (bottom).
Here $Z_{\rm m}$ is shown for radial bins with particles (if no particle
is found in a bin, $Z_{\rm m}$ is not plotted for the bin).
}
\label{Figure. 13}
\end{figure}

\begin{figure}
\psfig{file=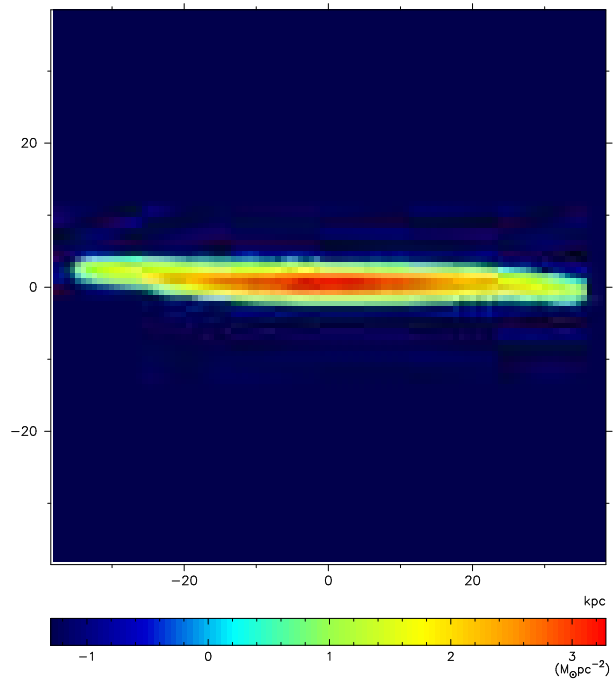,width=8.0cm}
\caption{
The same as Fig. 10 but for the $x$-$z$ projection in the model T18.
Owing to the short-term evolution ($\sim 1.4$ Gyr), a bulge-like
component in the EDC can not form in this model.
}
\label{Figure. 14}
\end{figure}

\subsection{Parameter dependences}

\subsubsection{Galactic precession}

(1) The time evolution of $\theta_{\rm d}$ and $\dot{ {\theta}_{\rm d} }$
depends on $M_{\rm dm, l}$,
$f_{\rm v,t}$ and $f_{\rm v, r}$,
because the orbital period and the last pericenter distance
of the LMC ($R_{\rm p}$) strongly depend on the three parameters. 
The present $\dot{ {\theta}_{\rm d} }$
is likely to be larger for models with larger $M_{\rm dm, l}$
for a given set of orbital parameters 
($f_{\rm v, t}$ and $f_{\rm v, r}$),  because
the tidal perturbation from the LMC is stronger thus
can more strongly influence the dynamical evolution of the Galaxy.
Both multiple passage and first passage  models 
show {\it negative} 
$\dot{ {\theta}_{\rm d} }$ 
in the present Galaxy. 
The Galaxy shows the maximum  $\theta_{\rm d}$  at the pericenter passage
of the LMC
and then $\theta_{\rm d}$ decreases. Therefore, the present Galaxy,
which is just after the pericenter passage of the LMC,
can show {\it negative}
$\dot{ {\theta}_{\rm d} }$.

(2) Although the present $\theta_{\rm d}$ is smaller
in first passage models than in multiple ones, 
$\dot{ {\theta}_{\rm d} }$ is larger in first passage models.
Fig. 12 shows the time evolution
of $\theta_{\rm d}$ for  the three  first passage models with
different $M_{\rm dm, l}$ yet the same $f_{\rm v, t}$ and $f_{\rm v, r}$. 
The present  $\theta_{\rm d}$ and
the present absolute magnitude of $\dot{ {\theta}_{\rm d} }$ 
are  larger for models with
larger  $M_{\rm dm, l}$ in the three models.

(3) The mean $\theta_{\rm d}$ and  $|\dot{ {\theta}_{\rm d} }|$ 
for these models are 0.64$^{\circ}$ and 4.6 degree Gyr$^{-1}$,
respectively.
These results strongly suggest  that the Galactic precession/nutation/pole-shift
due to
the last LMC-Galaxy tidal encounter can be detected by future observational
studies by GAIA, if there is a novel method to detect 
the Galactic precession/nutation/pole-shift.
We discuss future observational studies by GAIA for the detection of
the Galactic precession/nutation/pole-shift  later in \S 4.

\subsubsection{The influences on the EDC}

(1) The present outer structure of the EDC in the Galaxy
depends more strongly on $M_{\rm dm, l}$ rather than on $f_{\rm v, t}$
and $f_{\rm v, r}$ both in first passage and multiple passage models.
Fig. 13 shows that the warps of the EDCs can be more clearly seen
in first passage models with larger $M_{\rm dm, l}$ owing 
to stronger tidal perturbation of the LMC.
Although the models 
with $M_{\rm dm, l}=9 \times 10^{10} {\rm M}_{\odot}$ and
with $M_{\rm dm, l}=1.2 \times 10^{11} {\rm M}_{\odot}$ 
show warp-like structures in the outer regions of the EDCs,
the models with $M_{\rm dm, l}$ less than $6 \times 10^{10} {\rm M}_{\odot}$
do not clearly show such structures. 
Fig. 14 shows that the EDC in the first passage model
with  $M_{\rm dm, l}=9.0 \times 10^{11} {\rm M}_{\odot}$
(T18) has a warp-like structure, in particular, in the outer
regions with negative $x$.

(2) The outer  region of the EDC
in the model with lower $M_{\rm dm}$ 
(T19 with a lower mass of the Galactic dark
matter halo) can be  too severely damaged by the LMC-Galaxy tidal
interaction. 
This result implies that if the LMC has an original mass
of $\sim 10^{11} {\rm M}_{\odot}$, the Galaxy should already have 
enough a large mass ($\sim 10^{12} {\rm M}_{\odot}$) about
several Gyr ago to avoid 
too much damage of its outer part due to the LMC-Galaxy interaction. 
Equally, if the Galaxy has a significantly 
smaller mass when the LMC enters into
the Galaxy for the first time, then the initial mass of the LMC
should be small enough to suppress the destruction of the EDC of
the Galaxy.

(3) The present structures of the EDCs depend also on the initial size
($R_{\rm edc}$) such that the dynamical influences of the past LMC-Galaxy
interaction can be less clearly seen in the models with smaller $R_{\rm eds}$.
For example, warp-like structures can not be so clearly seen in the
model T16 in comparison with the model T17 (and T1).
Whether the simulated warps look like an ``integral symbol'' or
  a ``bow'' depends on the viewing angles of observers.

\section{Discussion}

\begin{figure}
\psfig{file=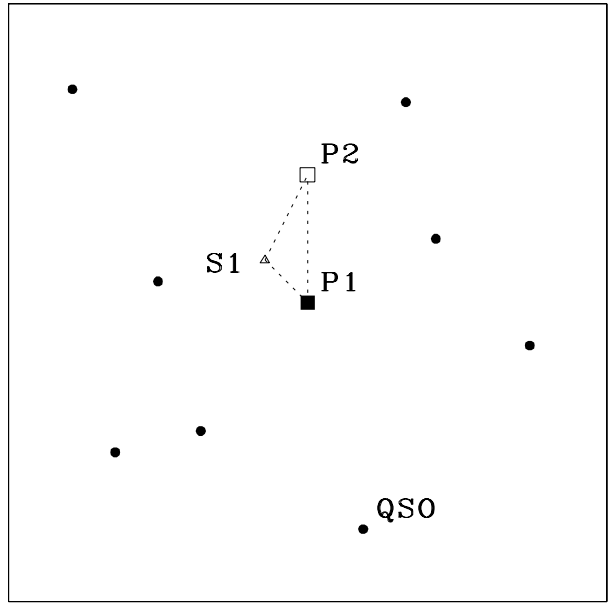,width=8.0cm}
\caption{
An illustrative figure for describing the proposed new method
by which we can possibly detect the Galactic precession. The QSOs
(or extragalactic radio sources) 
are shown by filled circles, and the locations of 
a hypothetical ``fixed point'' (e.g., the Sgr A$^{\star}$)
in the two observations done at two different epochs are denoted as
P1 and P2. The fixed point
(e.g., Sgr A$^{\star}$)  moves from P1 to P2 owing to (i) the 
Galactic precession (from P1 to S1) and (ii) the solar motion with respect
to the Galactic center (from S1 to P2).
}
\label{Figure. 15}
\end{figure}

\subsection{The Galactic precession}

The present study has first demonstrated that the LMC-Galaxy
interaction can induce the irregular precession/nutation/pole-shift
of the Galaxy.
From now on in this section, we simply use the term ``precession''
to describes the irregular precession/nutation or pole shift just
for convenience.
The mean value of the simulated
present precession  rates ($\dot{ {\theta}_{\rm d} }$)  is about
4.6 degree Gyr$^{-1}$
corresponding to 16.6 $\mu$as yr$^{-1}$. 
The derived $\dot{ {\theta}_{\rm d} }$ can be detected by future astrometry
satellites with $\mu$ac precision like GAIA, though there should be a novel way
to investigate the rate and the direction of the Galactic precession.
The LMC-Galaxy interaction would not be a sole mechanism for the induction
of the Galactic precession: minor merging (e.g., Quinn \& Goodman 1986) and
gas accretion onto the Galactic disk would also cause the Galactic 
precession.  The predicted precession direction and 
rate would be different between different
mechanisms,  future observations would be able to give strong constraints
on which mechanism would be the most reasonable for the possible precession.
Below we briefly discuss the implications of the ongoing Galactic precession
and a possible way to detect the precession.

\subsubsection{Implications}

If the Galactic precession is really ongoing, then it has a number of important implications.
First, changes in the $x$-, $y$-, and $z$-positions of an astronomical object
(outside the Galactic disk) in
the Galactic coordinate system for a certain period  of time
can be due to the combination of (i) the real change
of the 3D position with respect to the Galactic center
and (ii) the slight change in the orientation of the rotating axis of the Galactic
disk.
However,  the latter effect (ii)
would be significantly smaller  (e.g., $\dot{{\theta}_{\rm d}} \approx 4.6 \times 10^{-9}$ degrees
yr$^{-1}$ or $16.6 \mu$as yr$^{-1}$
for the possible LMC infall) in comparison with the former (i)
for most astronomical objects close to the Galaxy.
Second, the inclination angels of orbital planes
of the outer Galactic satellite galaxies with respect to the Galactic plane
can be different between past and present
because of the recent tilting of the disk
(not because of their orbital changes due to dynamical friction etc),
if the satellite galaxies were accreted onto the Galaxy
well before previous group infall.

Third, reconstruction of the past orbit of a Galactic satellite galaxy using
N-body simulations
should be done more carefully.
Almost all previous models to reproduce the observed stellar and gaseous streams formed from
tidal destruction of Galactic satellite galaxies (e.g., the Sgr dwarf  and the SMC) 
assumed a {\it fixed} Galactic potential (e.g., GN96;
 Law \& Majewsky 2010).
However, if the Galactic precession has been ongoing until recently, as suggested
by the present study, then the adopted  assumption of the fixed Galactic potential can
be no longer valid: the time-dependent disk potential due to the tilting disk
would need to be considered,
in particular, for those orbiting close  to the disk.

\subsubsection{How can we detect ?}

If the Galactic precession is real, then locations of distant QSOs
and extragalactic radio sources 
can not be used as (background) fixed points 
{\it in the Galactic coordinate}
to measure the proper motions of
the Galactic stars
in a very precise way: the locations of distant QSOs in the Galactic coordinate
system can change due to the precession.
Owing to the location of the Sun
with respect to the Galactic center and its move within the Galaxy,
the changes in the locations of QSOs in the Galactic coordinate are
due to
(i) the Galactic precession  and (ii) the change of the Sun's location
with respect to the Galactic center: the Galactic coordinate system depends
on the location of the Sun (see Fig. 15 for an illustrative purpose).
Here we consider that it is reasonable to refrain from
using the Galactic coordinate
system  and instead to use distant  QSOs as fixed
points on the sky.

We thus propose the following  method to detect a possible Galactic
precession  by using the background QSOs.
In this method, the QSOs are assumed to be fixed points on the sky 
and the location of 
the exact center of the Galaxy (the  radio source Sagittarius  A$^{\star}$,
Sgr A$^{\star}$,
denoted as ``P1'' and ``P2'' in Fig. 15)
is investigated at different two observations
done  at two different epochs  with a certain time interval (e.g., two years).
The position of Sgr A$^{\star}$ on the sky in each of the two observations
is estimated with respect to nearby QSOs. 
The positional  change
of Sgr A$^{\star}$  (from P1 to P2) between the two observations
is due to (i) the Galactic precession (from P1 to S1) and 
(ii) the solar motion (from S1 to P2). 
Therefore,
if we can separately
estimate the contribution from the solar motion (i.e., a vector
connecting between S1 and P2),
then we can have some information on the Galactic precession
(i.e., a vector connecting between P1 and S1).

The potential problem of the above method is that it would be difficult
to derive the contribution of the Sun's motion:
we need to predict very precisely how the Sun moves with respect
to Sgr A$^{\star}$
to detect observationally the Galactic precession. 
Recently Reid \& Brunthaler (2004) have investigated
the proper motion of Sgr A$^{\star}$ with respect to two extragalactic
radio sources using VLBI and found that the apparent proper motion
of Sgr A$^{\star}$ relative to J1745-283 is 
$6.379 \pm 0.024$ mas yr$^{-1}$. They claimed that
(i) the proper motion can be largely explained by the orbit of the Sun
around the Galaxy and (ii) the residual proper motion perpendicular
to the plane of the Galaxy is $-0.4 \pm 0.9$ km s$^{-1}$.
This residual could have some information on the Galactic precession,
though there could be some observational uncertainties in the solar motion
within the Galaxy.

Although Sgr A$^{\star}$ is considered to be 
a hypothetical fixed point in the above method,
other fixed points (if any) within  the Galaxy can be proposed.
There could be  other possible  methods to detect the Galactic 
precession, but we currently do not have any better ones.
Although Hipparchus discovered the evidence of the Earth's precession
more than 2000 years ago,
the possibility of the {\it ongoing}  Galactic precession (and pole shift)
and the likely impact of the precession in the modern astronomy
have  not been extensively
discussed by observational and theoretical studies so far.
Given the $\mu$as accuracy of future astrometry satellites,
now the time seems to be ripe for extensive discussion on the origin
of the possible Galactic precession and pole shift.

\begin{figure}
\psfig{file=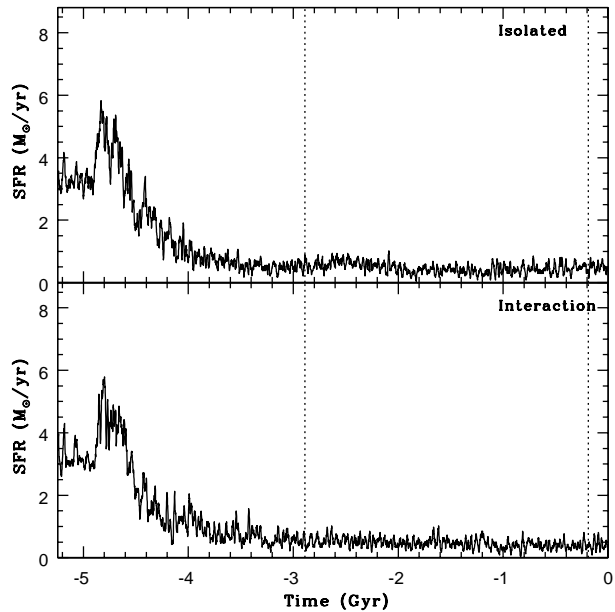,width=8.0cm}
\caption{
The time evolution of star formation rates in the isolated model
(upper) and the interaction one (lower). The epochs of the two pericenter
passages of the LMC are shown by dotted lines.
}
\label{Figure. 16}
\end{figure}

\begin{figure}
\psfig{file=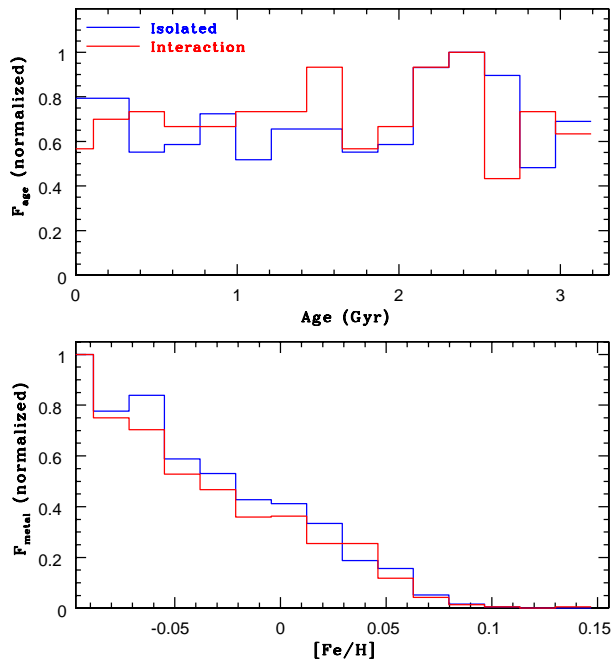,width=8.0cm}
\caption{
The age (upper) and metallicity (lower) distributions of new stars 
formed from gas in the Galaxy for the isolated (blue) and interaction
(red) models. The distributions are normalized by the maximum number
of stars in the 20 age and metallicity bins. 
}
\label{Figure. 17}
\end{figure}

\subsection{The Galactic star formation history influenced
by the LMC ?}

Rocha-Pinto et al. (2000) investigated the age distribution of
552 late-type dwarfs and thereby
derived the star formation history of the Galaxy.
They showed that the disk of the Galaxy has experienced  enhanced episodes
of star formation at $0-1$ Gyr, $2-5$ Gyr, and $7-9$ Gyr ago and suggested
that these enhanced episodes can be closely associated with the interaction
between the Magellanic Clouds and the Galaxy.
There could be a number of physical mechanisms responsible
for the observed possible enhancement of the Galactic star formation:
for example, minor merging and rapid infall of high velocity clouds
could enhance significantly star formation in some local regions of 
the Galaxy.
However, we here  
focus exclusively on whether the LMC-Galaxy tidal interaction can significantly
change the star formation history of the Galaxy. 

In order to discuss the possible influence of the LMC-Galaxy interaction
on the Galactic star formation history,
we use chemodynamical 
numerical simulations with star formation  and chemical evolution.  
Since our GRAPE-based and GPU-based numerical codes are only for 
collisionless system, we adopt our TREESPH code used in our previous
numerical simulations on the formation of stars and globular clusters
in interacting galaxies (see Bekki et al. 2002
for definitions and details).  We do not include the
model for globular cluster formation (that was modeled
in Bekki et al. 2002) in the present investigation.
We adopt the present low-resolution model and add 
a gas disk with an exponential radial profile  to
the Galaxy so that star formation from gas can be investigated.
The gas disk has a size of $2R_{\rm d}$,
a scale length of $0.4R_{\rm d}$, 
and a scale hight of $0.02R_{\rm d}$, and a mass of $0.2 M_{\rm d}$.
Initially the gas disk has a metallicity of 0.016 and the chemical yield
and the return parameter in the chemodynamical model are set to be
0.01 and 0.3, respectively. An isothermal equation of state 
is adopted for the gas with a temperature of $10^4$ K.

Fig. 16 shows the star formation history of the Galaxy in the model
in which the model parameters are the same as those used in the standard
model. For comparison, the star formation history in the isolated model
(i.e., without interaction with the LMC) is shown.  Clearly, there is no
dramatic difference in the star formation history of the Galaxy between
the two models, which suggests that the LMC can not change significantly
the {\it global star formation history} of the Galaxy
(e.g., the mean star formation
rate for the last several Gyr) through its tidal interaction with
the Galaxy.
The star formation rate of the Galaxy can not be enhanced by the last LMC-Galaxy
interaction about 0.2 Gyr ago, though the LMC can strongly influence
the outer part of the Galactic disk.
There appears to be a small  difference in the Galactic star formation
history $\sim 1$ Gyr after (i.e., $\sim 1.6$ Gyr ago)
the first pericenter passage of the LMC
between the two models: the LMC's tidal force could possibly  have a very minor  
influence on the 
Galactic star formation. But the difference is as small as the noise
fluctuation, and thus we can not claim that the LMC can increase 
global star formation rate of the Galaxy  in the present study.

Fig. 17 shows the present age and metallicity distributions of new stars formed 
from gas for the disk region with 7.0 kpc $\le R \le$ 10 kpc 
(i.e., the solar-neighborhood) 
in the isolated and interaction models.
Although the differences in the age distribution between the 
isolated and interaction models can be barely seen
(the difference can be as small as the noise fluctuation), 
there is no outstandingly clear peak at an age of $\sim 3$ Gyr
when the first LMC-Galaxy encounter can possibly  influence the star formation
history in the solar-neighborhood.
Furthermore there is no clear peak 
in the age distribution of the interaction model at an age of $0.1-0.2$ Gyr
where the last LMC-Galaxy tidal encounter occurs.
These results imply that it is a formidable task for observational
studies to find the possible evidence 
(e.g.,  a burst epoch of star formation) 
for the Galactic star formation
in the solar-neighborhood influenced by
the LMC-Galaxy interaction.
As shown in Fig. 17, there are some differences
in the metallicity distributions of stars in the solar-neighborhood
between the isolated and interaction models.
Although the differences enable us to claim that the LMC-Galaxy
interaction can slightly change the metallicity distribution of the stars,
they do not provide strong  constraints on when the LMC-Galaxy interaction
strongly influenced the star formation in the solar-neighborhood.

\begin{figure}
\psfig{file=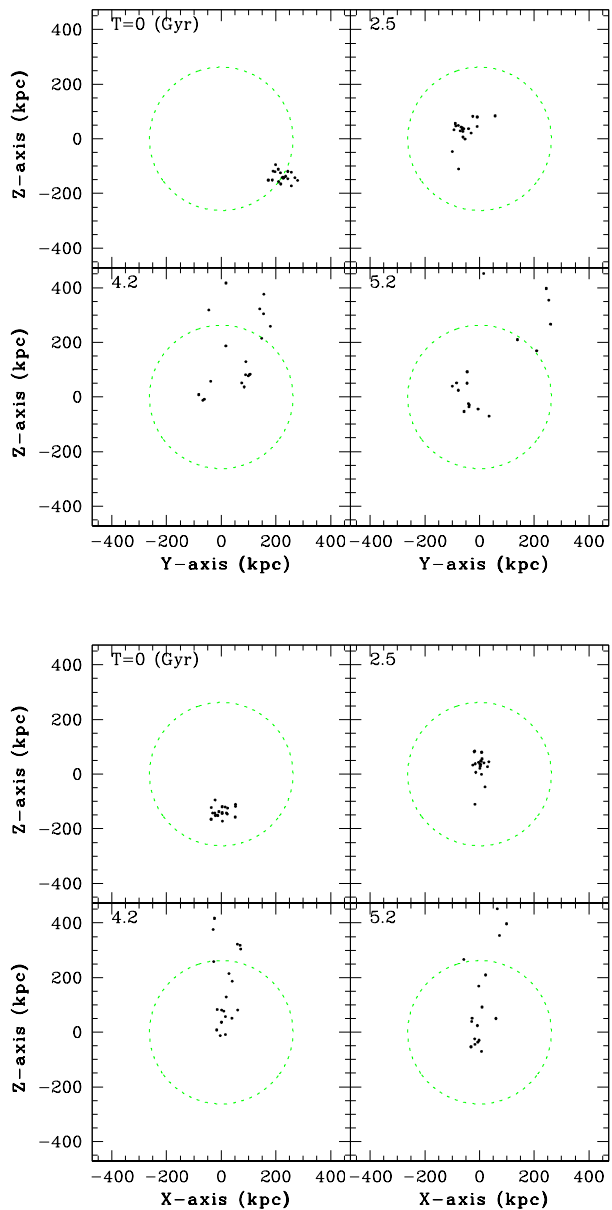,width=8.0cm}
\caption{
The time evolution of the distribution of the 20 dwarfs initially
in the LMC group projected onto the $y$-$z$ plane (upper four)
and the $x$-$y$ one (lower). The virial radius of the Galaxy
is shown by a green dashed line in each panel.
}
\label{Figure. 18}
\end{figure}

\subsection{The distribution of the stripped satellite galaxies from the
LMC group}

So far we have described the results of the models in which dwarfs
are not included, though the LMC could have been accreted onto the 
Galaxy as a group with a number of dwarfs (e.g., Bekki 2008).
The reason for this  non-inclusion is that small galaxy groups are dominated
by massive dark matter halos (i.e., the total masses of group member dwarfs
are quite small) so that the orbital evolution of the groups can be determined
largely by the total masses of dark matter initially in the groups.
Indeed, even if 20 dwarfs with the total mass
being  0.1\% of the total mass of the dark matter in a group
are added to the group, the present results 
do not change at all.  The models with 20 equal-mass dwarfs 
(i.e., the models T23 and 24) are useful
in discussing how the infall of the LMC with dwarfs can change the
distribution of the Galactic satellite galaxies.

One of interesting results in the model T23  with 20 dwarfs is that the
group member dwarfs show a thin disk-like distribution 
along the $z$-axis within $R<400$ kpc  in the Galaxy.
As shown in Figure 11,  the thin distribution of the group
member dwarfs (some of which are tidally stripped during group infall)
is closely associated with the dark disk formed during group infall.
This result implies that the observed thin disk of the Galactic satellite
galaxies along the Galactic polar axis (e.g., Metz et al. 2009)
can be understood in the context of
a previous group infall event of the LMC 
around the Galaxy. The final distribution of the dwarfs stripped from
the LMC depends on $R_{\rm dw}$ such that the thin distribution
of the stripped dwarfs can be more clearly seen in the model
with larger $R_{\rm dw}$ (T23 compared to T24).Recently Nichols et al. (2011)
have shown that Draco, Sculptor, Sextans, Ursa Minor, and the Sagittarius
Stream are consistent with them initially in the LMC group,
though their models are rather idealized (using test particle simulations).

It should be noted here that two dwarfs can sink into the inner region
of the LMC disk ($R<R_{\rm d, l}$) owing to dynamical friction against
the halo in the model T24 with $R_{\rm dw}=5R_{\rm d, l}$:
these dwarfs can be regarded as being merged with the LMC.
This  possible merging between the LMC and the group member dwarfs does not
occur in the model T23 with  $R_{\rm dw}=10R_{\rm d, l}$. 
Although a  possible minor merger event in the LMC was discussed in
the context of the observed possible counter-rotating component of stars
in the LMC (e.g., Subramaniam \& Prabhu 2005), 
its influence on the LMC's evolution 
(e.g., the thick disk formation) has not been
discussed extensively in previous theoretical studies.
It is thus our future study how such a minor merger event can change
the star formation history and dynamical evolution of the LMC.

\section{Conclusions}

We have developed a new method by which we can reproduce 
rather precisely and efficiently  the present
location of the LMC  in self-consistent N-body numerical
simulations of orbital evolution of the LMC  
in a live gravitational  potential of the Galaxy.
Using this new method (called ``CTM''),
we have investigated the orbital evolution of the LMC from outside
the virial radius of the Galaxy and its influence on the evolution of 
the Galaxy in a self-consistent manner for a given set of model
parameters of the LMC and the Galaxy. The principle results are 
summarized as follows.

(1) The LMC-Galaxy tidal interaction can cause the 
``pole shift'' (or irrugular precession/nutation)   of the Galaxy
with the present pole-shift
rate $\dot{ {\theta}_{\rm d} }$ being $\sim 2$ degree Gyr$^{-1}$
(corresponding to  $\sim 7 \mu$as yr$^{-1}$), if the original mass of the
LMC before its entrance into the virial radius of the Galaxy is as large
as $9 \times 10^{10} {\rm M}_{\odot}$.  The values of the present
$\dot{ {\theta}_{\rm d} }$ depend on $M_{\rm dm, l}$ and orbits
of the LMC, and  the mean $\dot{ {\theta}_{\rm d} }$ for models
that reproduce the present location of the LMC
is $\sim 4.6$ degree Gyr$^{-1}$
(corresponding to 16.6 $\mu$as yr$^{-1}$).

(2) The outer part ($R>20$ kpc)  of the extended disk
component (EDC) of the Galaxy can be 
influenced by the LMC-Galaxy tidal interaction, in particular, when the
LMC passes through its last pericenter of the orbit. As a results of this,
the outer part of the EDC shows warp-like structures in most models.
The detailed morphologies of the warps (e.g., ``integral symbol''
or ``bow-like'') depend on the viewing angles of observations
and $M_{\rm dm, l}$. The preset study thus confirms the results of 
previous theoretical studies on the formation of the Galactic warp
by the LMC-Galaxy interaction.

(3) The LMC-Galaxy tidal interaction can also be responsible for the formation
of  elongated and asymmetric (e.g., spiral-like) structures in the outer
part of the EDC. The interaction can  dynamically heat up
the EDCs so that the vertical velocity dispersions of the outer
parts ($R>20$ kpc) of the EDCs in the tidal models
can be significantly larger than those in the isolated models.
The structure and kinematics  of the EDC depend on the original mass
and the orbit of the LMC. 
The outer part of the Galaxy can thus have fossil records
of the past LMC-Galaxy tidal interaction.
The LMC-Galaxy interaction can change
the outer structure and kinematics of the Galaxy also in first passage
models.

(4) The global star formation history of the Galaxy for the last several
Gyr (e.g., the mean star formation rate across the Galaxy) can not be 
significantly influenced by the LMC-Galaxy dynamical interaction,
if the initial $M_{\rm dm, l}$ is less than 
$1.2 \times 10^{11} {\rm M}_{\odot}$.
The star formation history
and the age-metallicity relation of stars at the solar-neighborhood
($7<R<10$ kpc defined in the present study) can be only slightly
influenced by the LMC-Galaxy interaction. However, these results need
to be re-investigated and confirmed by future more sophisticated and
higher resolution numerical simulations.

(5) If the LMC
is accreted onto the Galaxy as group with dwarfs,
then the stripped dwarfs during the group infall can show a thin disk-like
distribution along the $z$-axis
within $R<400$ kpc around the Galaxy. The distribution of the dwarfs
follows the disky/ring-like distribution of dark matter stripped from the group.
Therefore,  both the possible Galaxy precession/nutation/pole-shift
and the observed thin distribution of the Galactic satellite
galaxies can share a common origin (i.e., the infall of the LMC  group).
A few of the dwarfs in the group can merge with the LMC during the
infall of the LMC, if the initial distribution of the dwarf is more compact.

(6) The derived  $\dot{ {\theta}_{\rm d} }$ 
of the Galaxy   caused
by the LMC infall is suggested to be able to be
detected by future astrometric satellites
with $\mu$as precision (such as GAIA).
The possible Galactic precession 
has a number of important implications in galactic astronomy,
such as the observational derivation of the proper motions of Galactic satellite
galaxies using the background QSOs.
A possible way to detect the ongoing precession of the Galaxy
has been briefly discussed.

(7) We conclude that the dynamical evolution of the outer part ($>20$ kpc)
of the Galaxy can be influenced by the past long-term LMC-Galaxy
tidal interaction.  
Therefore the structure and kinematics of the outer part of the Galaxy
can have fossil records  on  when and how the LMC arrived in the Galaxy.
It is doubtlessly worthwhile for future observational studies to investigate
whether there is an imprint of the past LMC-Galaxy interaction in
the star formation history of the solar-neighborhood.

\section{Acknowledgment}
I am   grateful to the anonymous referee for constructive and
useful comments that improved this paper.
KB acknowledge the financial support of the Australian Research Council
throughout the course of this work.
The work was supported by iVEC through the use of advanced computing resources located at the University of Western Australia.

\appendix

\section{The orbital evolution of the LMC in unsuccessful
models}

\begin{figure}
\psfig{file=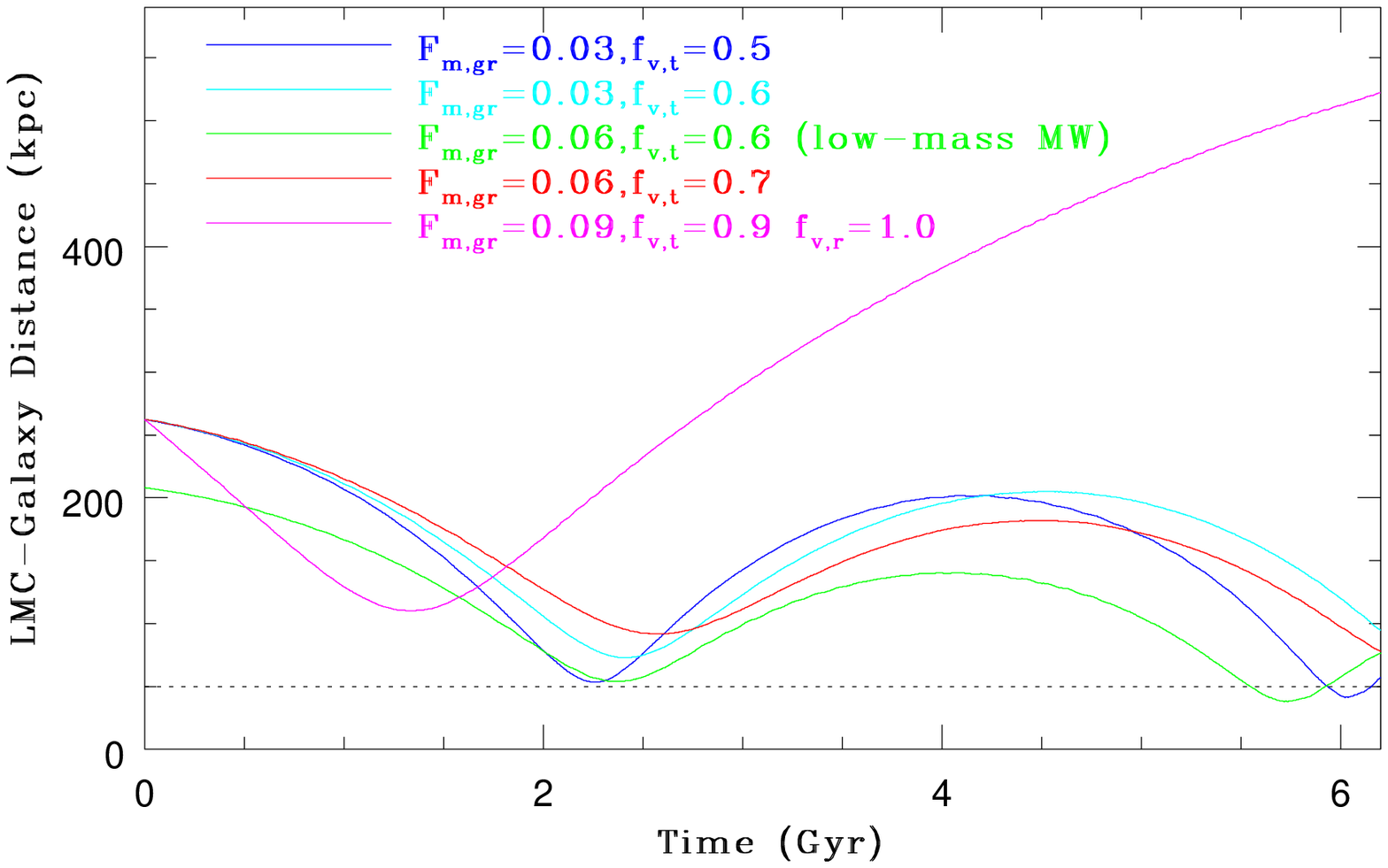,width=8.0cm}
\caption{
The orbital evolution of the LMC in
the five models:
$F_{\rm m, gr}=0.03$ and $f_{\rm v,t}=0.5$ (blue),
$F_{\rm m, gr}=0.03$ and $f_{\rm v,t}=0.6$ (cyan),
$F_{\rm m, gr}=0.06$ and $f_{\rm v,t}=0.6$ (green),
$F_{\rm m, gr}=0.06$ and $f_{\rm v,t}=0.7$ (red),
and $F_{\rm m, gr}=0.09$, $f_{\rm v,t}=0.9$, and $f_{\rm v,r}=1.0$ (magenta).
For the models with 
$F_{\rm m, gr}=0.03$ and 0.06,  $f_{\rm v, r}=0.2$. 
The dark matter mass of the Galaxy in
the model shown by a green line
($F_{\rm m, gr}=0.06$ and $f_{\rm v,t}=0.6$) is
$5 \times 10^{11} {\rm M}_{\odot}$ (i.e., the half of $M_{\rm dm, mw}$
in other models).
The present LMC-Galaxy
distance is shown by a dotted line.
}
\label{Figure. 19}
\end{figure}

In the main text,
we have mainly described the results on the ``successful models''
in which the observed
position of the LMC can be reproduced well.  
The pericenter radius ($R_{\rm p}$) of the LMC's orbit around the Galaxy
can not be as small as $R=50$ kpc within $\sim 8$ Gyr 
in the ``unsuccessful models''
with lower $M_{\rm dm, l}$ and higher $f_{\rm v}$.
This is because the time scale of dynamical friction
of the LMC can be long (longer than the Hubble time) owing to
the smaller LMC masses and the larger amount of initial
orbital angular momentum in these models.
 Although these models
are not useful in discussing the dynamical influences of the LMC on the Galaxy,
they are quite useful in determining a range of $f_{\rm v}$ 
($f_{\rm v, t}$ and $f_{\rm v, r}$) for realistic orbital models
of the LMC and thus discussing how the LMC
was accreted onto the Galaxy.
Fig. A1 summarizes the results of five  different models, three of which are
regarded as unsuccessful models in the present study.
The mass-ratio of $M_{\rm dm, l}$ to $M_{\rm dm, mw}$ is referred to
as $F_{\rm m, gr}$ for convenience and used in this and following appendix
sections.

As shown in Fig. A1,
the LMC can not have $R_{\rm p} \le 50$ kpc within $\sim 8$ Gyr in
the model T9 with $M_{\rm dm, l}=3 \times 10^{10} {\rm M}_{\odot}$ 
($F_{\rm m, gr}=0.03$)
owing to inefficient dynamical friction of the less massive LMC.
This result implies that
if the Galaxy was already as massive as $M_{\rm dm, mw}=10^{12} {\rm M}_{\odot}$
at the epoch of the LMC accretion onto the Galaxy,
then the LMC should have a mass significantly larger than
the present mass ([$1-2] \times 10^{10} {\rm M}_{\odot}$) at its accretion epoch
in order to reach $R=50$ kpc in the subsequent orbital evolution. 
If the  Galaxy has a lower mass of $5 \times 10^{11} {\rm M}_{\odot}$,
then the LMC with $M_{\rm dm, l}=3 \times 10^{10} {\rm M}_{\odot}$
can show $R_{\rm p} \le 50$ kpc in the model T21.
This result implies that if the LMC
has a lower original mass 
($M_{\rm dm, l} \sim 3 \times 10^{10} {\rm M}_{\odot}$),
then the Galaxy should have a lower mass at the epoch of the LMC accretion
for the LMC to finally reach $R=50$ kpc.
Therefore the results of the models T9 and T21 suggest that
the present location and orbit of the LMC can give constraints on the
original mass of the LMC and the Galaxy mass at the LMC accretion.

The LMC in the model T14 with 
$M_{\rm dm, l}=6 \times 10^{11} {\rm M}_{\odot}$
($F_{\rm m, gr}=0.06$),
$f_{\rm v, t}=0.7$,
and $f_{\rm v, r}=0.2$ can not reach $R=50$ kpc within $\sim 8$ Gyr,
though the LMC in the model T4 with  
$M_{\rm dm, l}=9 \times 10^{11} {\rm M}_{\odot}$
($F_{\rm m, gr}=0.09$)
and the same $f_{\rm v, t}$
and $f_{\rm v, r}$ can reach $R=50$ kpc within  $\sim 6$ Gyr.
This result suggests that the LMC with a lower mass needs  to have
lower $f_{\rm v, t}$ (for a fixed $f_{\rm v, r}$) to reach 
$R=50$ kpc within $\sim 8$ Gyr. The models with higher $f_{\rm v, t}$ ($>0.7$)
and higher $f_{\rm v, r}$ ($>1$) do not show $R_{\rm p} \le 50$ kpc for 
the LMC. For example, the orbit of the LMC shows the 
pericenter of more than 100kpc
and the apocenter much larger than the virial radius of the Galaxy
in the model T8 with
$f_{\rm v, t}=0.9$
and $f_{\rm v, r}=1.0$.
These results in Fig. A1 clearly demonstrate that there is a required range
of model parameters for the LMC to finally reach $R=50$ kpc since the LMC 
accretion onto the Galaxy from outside of the Galactic virial radius.

\section{The Galactic precession due to accretion of small groups
and satellite galaxies}

\begin{figure}
\psfig{file=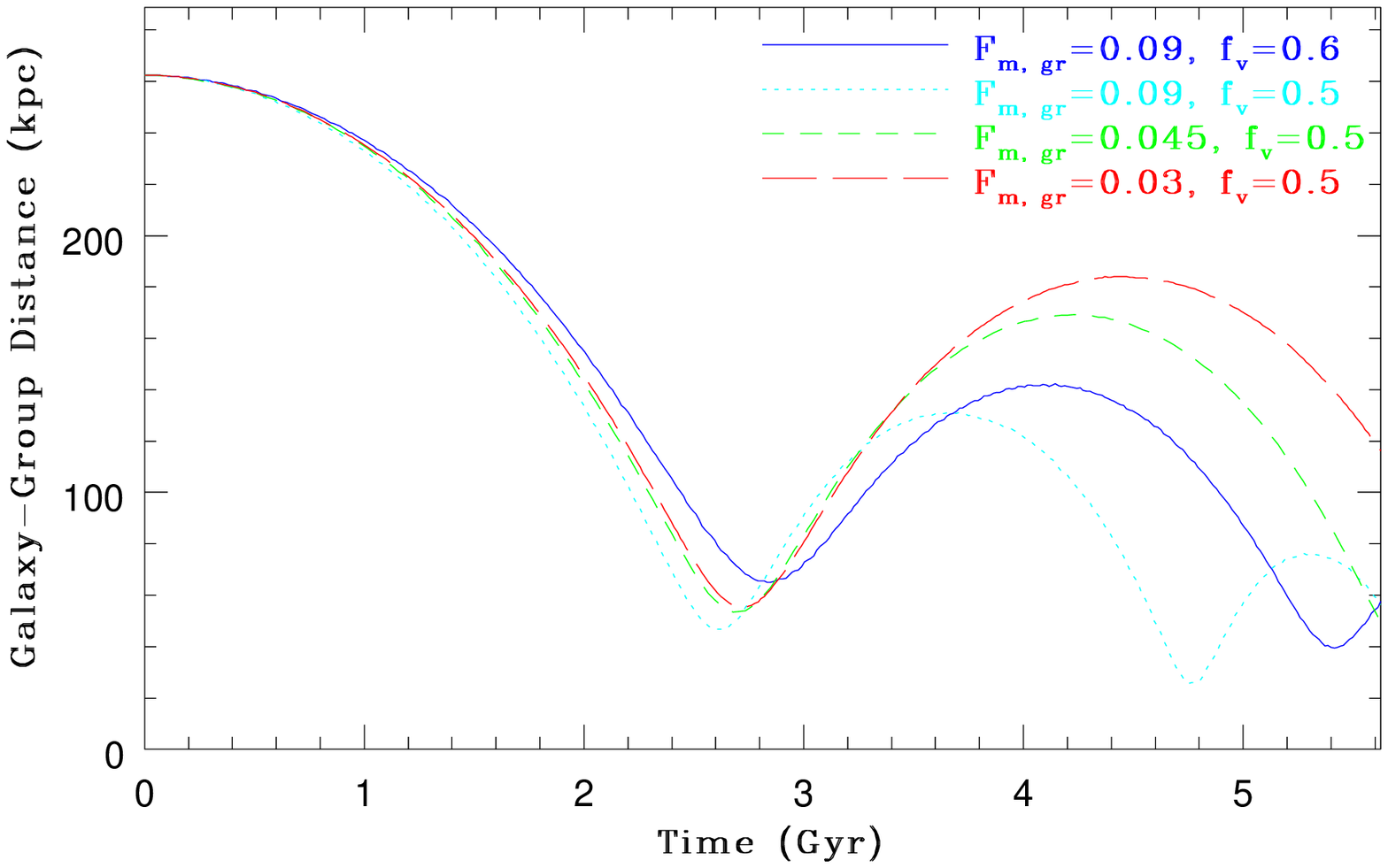,width=8.0cm}
\caption{
The time evolution of the distance  between a disk galaxy and a group infalling onto
the galaxy for the last 5.6 Gyr in four different models with
$F_{\rm m, gr}=0.09$ and $f_{\rm v}=0.6$ (blue solid),
$F_{\rm m, gr}=0.09$ and $f_{\rm v}=0.5$ (cyan dotted),
$F_{\rm m, gr}=0.045$ and $f_{\rm v}=0.5$ (green short-dashed),
and $F_{\rm m, gr}=0.03$ and $f_{\rm v}=0.5$ (red long-dashed).
For these models, ${\theta}=60^{\circ}$.
}
\label{Figure. 20}
\end{figure}

\begin{figure}
\psfig{file=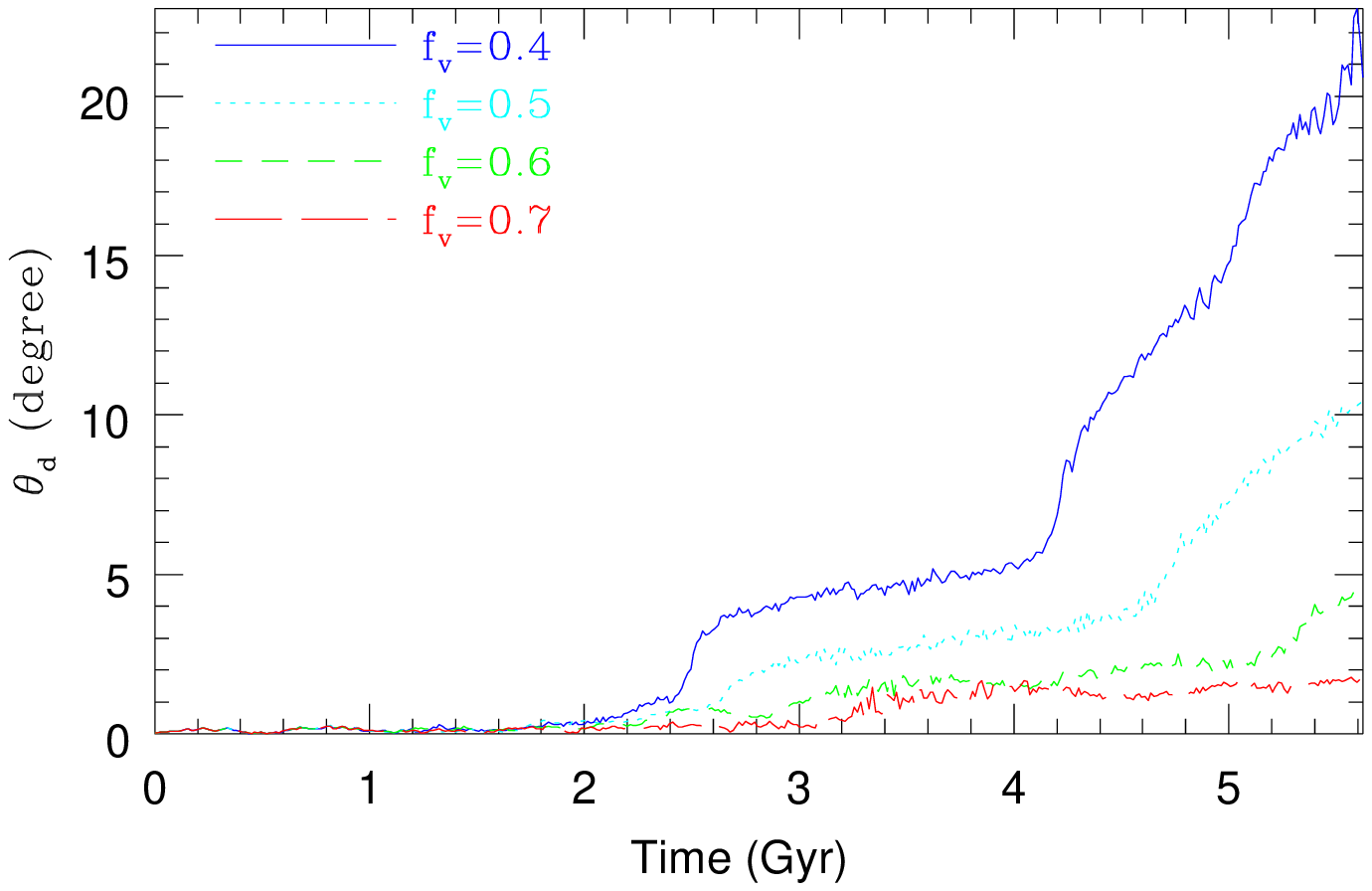,width=8.0cm}
\caption{
The time evolution of the angle (${\theta}_{\rm d}$) between the $z$-axis
and the orientation of the rotation axis of a stellar disk
in four different models with
$f_{\rm v}=0.4$ (blue solid),
$f_{\rm v}=0.5$ (cyan dotted),
$f_{\rm v}=0.6$ (green short-dashed),
and $f_{\rm v}=0.7$ (red long-dashed).
For these models,  $F_{\rm m, gr}=0.09$ and ${\theta}_{\rm gr}=60^{\circ}$.
}
\label{Figure. 21}
\end{figure}

\begin{figure}
\psfig{file=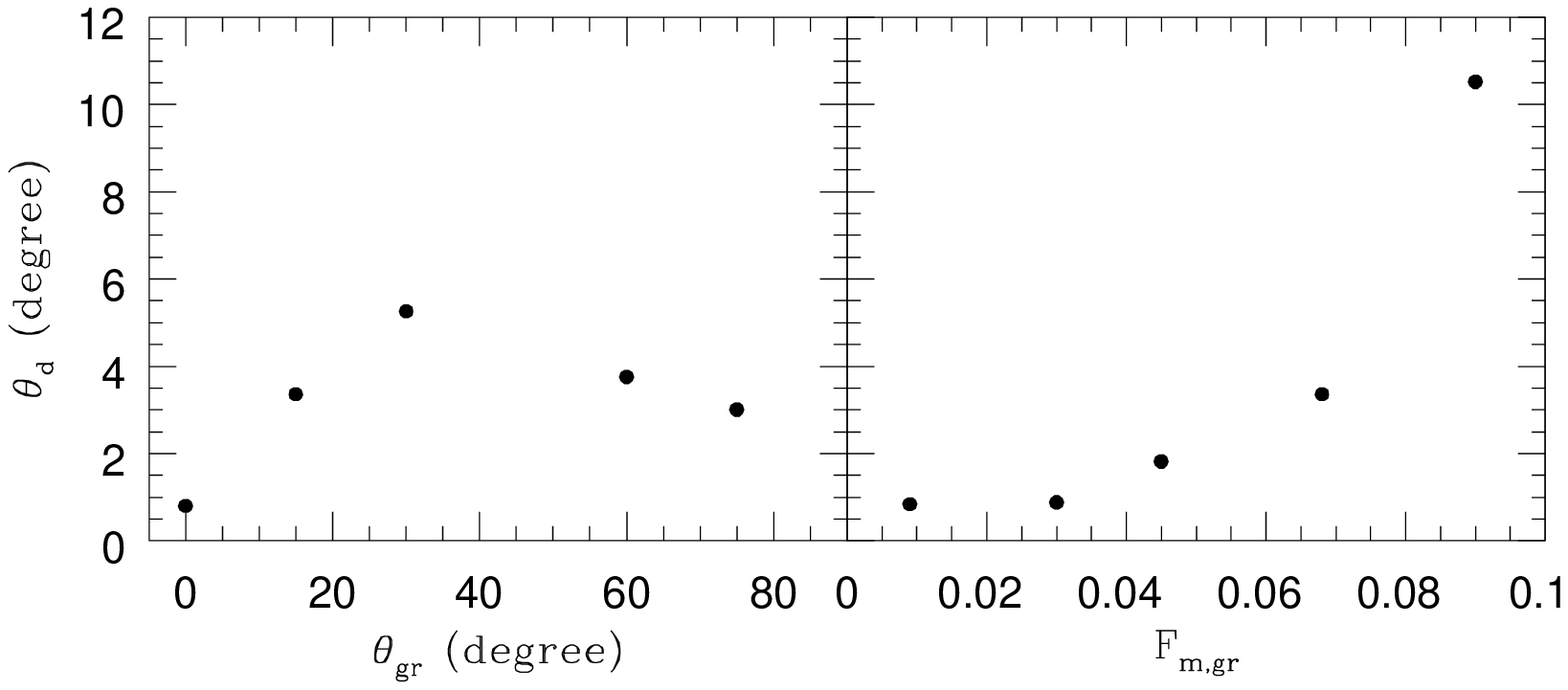,width=8.0cm}
\caption{
The dependences of ${\theta}_{\rm d}$ on ${\theta}_{\rm gr}$ (left)
and $F_{\rm m, gr}$ (right).
The  models in the left panel have $F_{\rm m, gr}=0.09$ and $f_{\rm v}=0.6$
whereas those in the right panel have $f_{\rm v}=0.5$ and ${\theta}_{\rm gr}=60^{\circ}$.
}
\label{Figure. 22}
\end{figure}

\begin{figure}
\psfig{file=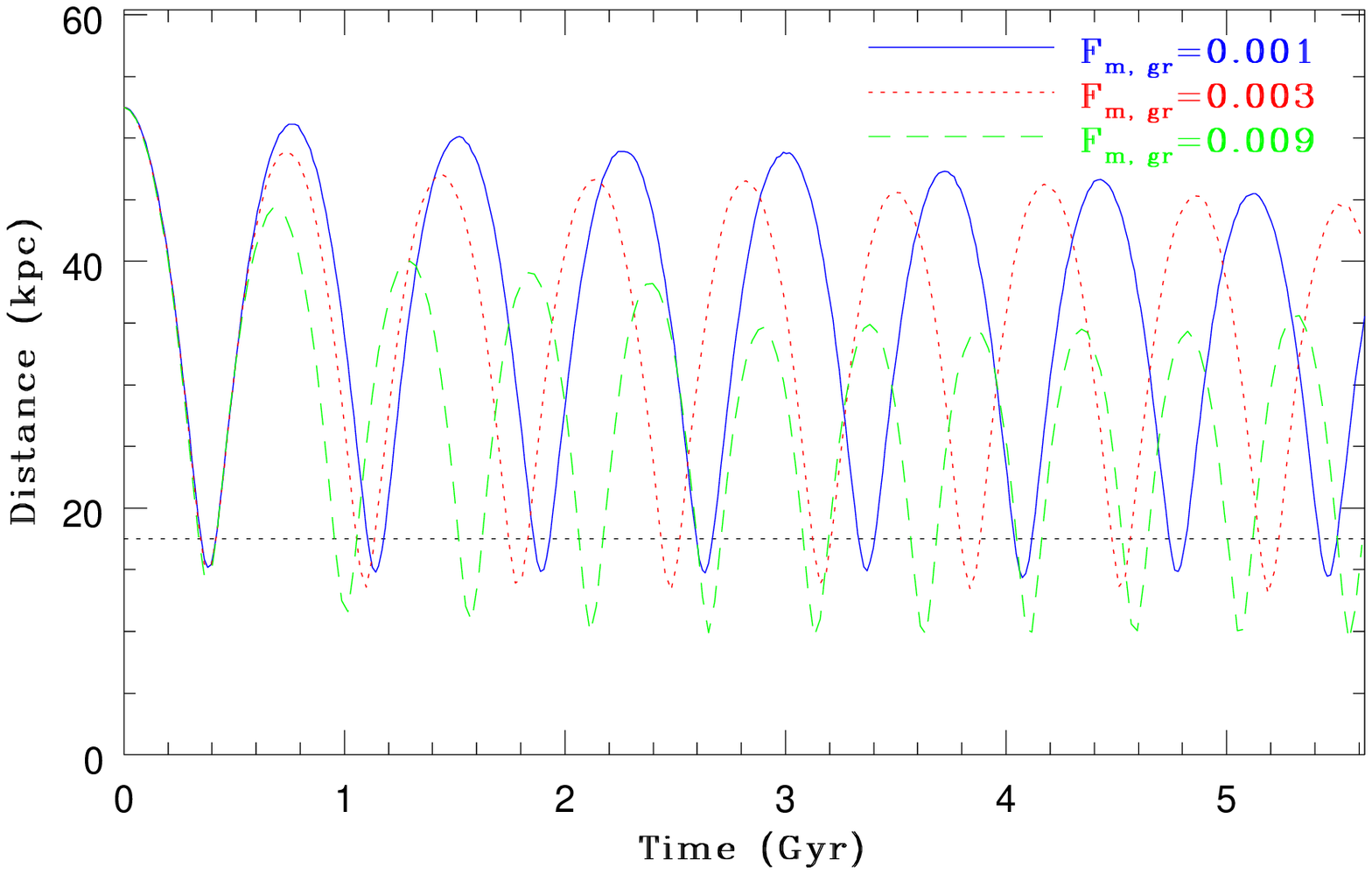,width=8.0cm}
\caption{
 The time evolution of distance between a disk galaxy and an infalling group
(including a dwarf) for three different models with $F_{\rm m, gr}=0.001$ (blue
solid), 0.003 (red dotted), and 0.009 (green short-dashed).
For these models, $f_{\rm v}=0.5$ and ${\theta}_{\rm gr}=60^{\circ}$.
The size of the stellar disk is shown
by a black dotted line for comparison.
}
\label{Figure. 23}
\end{figure}

\begin{figure}
\psfig{file=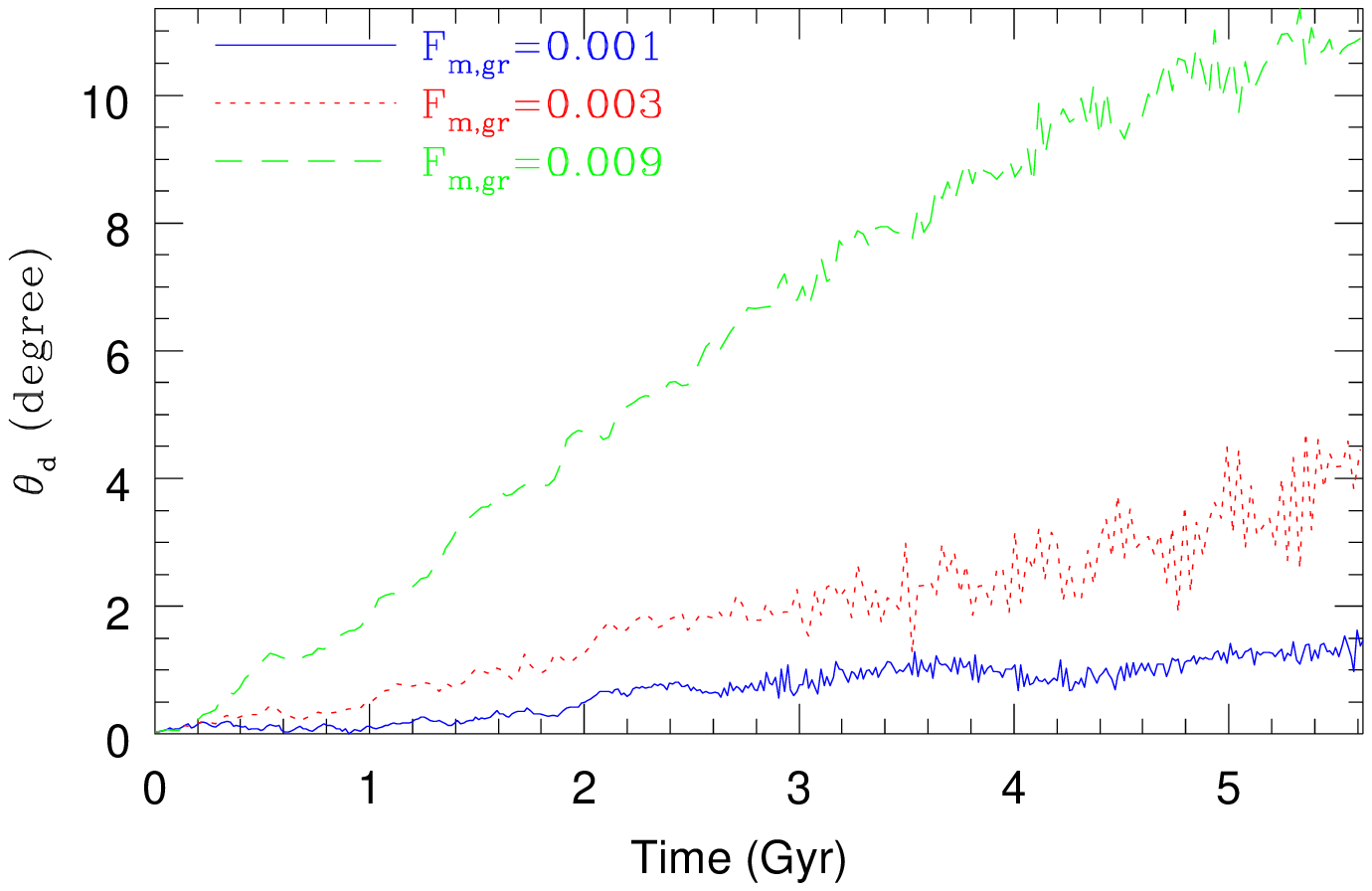,width=8.0cm}
\caption{
The time evolution of ${\theta}_{\rm d}$
of a  disk galaxy  for three different models
shown in Figure 8.
}
\label{Figure. 24}
\end{figure}

Recent high-resolution cosmological simulations of 
the formation of a Milky Way-like
halo in a $\Lambda$ cold dark matter cosmology have shown that
a significant fraction of the subhalos have been accreted 
in group (e.g., Li \& Helmi 2008). Since the orbital parameters of
the groups accreting onto the Galaxy can be significantly different
from those of the LMC (or the LMC group), it is worthwhile for the
present study to briefly discuss how group infall can influence the
dynamical evolution of the Galaxy for different orbital parameters
of the groups. Here we focus on how $\theta_{\rm d}$ depends on 
(i) $F_{\rm m, gr}$ (i.e., the mass ratio of a group to the Galaxy)
and (ii) $\theta_{\rm gr}$, which is the inclination angle between the 
orbital plane of a group and the $x$-$y$ plane.

Fig. B1 shows the orbital evolution of a group in the Galaxy 
for  models with different $F_{\rm m, gr}$ and
$f_{\rm v}$ for the last 5.6 Gyr. Owing to dynamical friction of the group
against the dark matter halo of the Galaxy,  
the pericenter distance ($R_{\rm p, gr}$)  of
the orbit becomes progressively smaller 
as the group moves around  the disk: $R_{\rm p, gr}$
for the first and second pericenter passages are 46.6 kpc and 33.7 kpc, respectively,
in the model with $F_{\rm m, gr}=0.09$ and $f_{\rm v}=0.5$. 
The final $R_{\rm p, gr}$ depends on $F_{\rm m, gr}$ and $f_{\rm v}$
such that (i) $R_{\rm p, gr}$ is larger in models with larger $f_{\rm v}$ 
for a given  $F_{\rm m, gr}$ and (ii) it is larger for smaller $F_{\rm m, gr}$
for a given $f_{\rm v}$.  Therefore the disk galaxy can  more strongly feel 
the time-changing tidal torque from the infalling group in models
with smaller $f_{\rm v}$ and larger $F_{\rm m, gr}$
so that the stellar disk can be strongly influenced 
by the group infall in these models.

Fig. B2 shows that ${\theta}_{\rm d}$ of the stellar disk in each of
the models with the same $F_{\rm m, gr}$ yet 
different $f_{\rm v}$  becomes larger as galaxy-group
interaction proceeds during group infall. The time evolution of ${\theta}_{\rm d}$
is more dramatic in the models with smaller $f_{\rm v}$ owing to
smaller $R_{\rm p, gr}$ thus stronger tidal torque of groups.
The final ${\theta}_{\rm d}$ in the models with $f_{\rm v}=$0.4, 0.5, 0.6, and 0.7
are 22.7$^{\circ}$, 10.5$^{\circ}$, 4.5$^{\circ}$, and 1.8$^{\circ}$, respectively:
galactic precession (pole shift)
 is more remarkable in the models with smaller $f_{\rm v}$.
The rate of the precession 
($\dot{{\theta}_{d}}$) in the last 1 Gyr ($T=4.6-5.6$ Gyr) is also larger in the models
with smaller $f_{\rm v}$: $\dot{{\theta}_{d}}$ in the models with $f_{\rm v}=0.4$,
0.5, 0.6, and 0.7 are 8.9, 6.0, 2.3, and 0.6 degrees Gyr$^{-1}$, respectively.
In comparison with almost  steady increase of ${\theta}_{\rm d}$ for the last 1 Gyr,
the time evolution of ${\phi}_{\rm d}$ is much less steady in all models
of the present study.
As discussed later, the higher rates in the models with $f_{\rm v} \le 0.6$
(an order of $10\mu$as yr$^{-1}$) can be detected by future astrometry
satellites with $\mu$as precision like GAIA.

As the present study has shown, if the {\it initial} total mass
of a satellite galaxy including the extended massive dark matter halo
is as large as 0.3\% of the total mass of its host galaxy
and if the pericenter
distance of its orbit around the host is as small as $R_{\rm d}$ (=disk size),
then the satellite can cause an appreciable amount of the galactic precession 
(${\theta}_{\rm d} \approx 4^{\circ}$ and
$\dot{ {\theta}_{\rm d}} \approx 1^{\circ}$ Gyr$^{-1}$).
If we assume that the total mass of the Galaxy is
$10^{12} M_{\odot}$ (which is consistent with
the observationally inferred value by Wilkinson \& Evans 1999),
then the total mass of Sgr ($M_{\rm Sgr}$)
and the pericenter distance of the Sgr's orbit ($R_{\rm p, Sgr}$)
need to be at least $3 \times 10^9 M_{\odot}$
and less than $\sim 18$ kpc, respectively,
for Sgr to cause the Galactic precession 
with the rate of $\dot{ {\theta}_{\rm d}} \approx 1^{\circ}$ Gyr$^{-1}$.

As shown in Fig. B3,
the comparative model with ${\theta}_{\rm gr}=0^{\circ}$ shows very small
final ${\theta}_{\rm d}$ ($0.8^{\circ}$),
which means  that the orbit of a group infalling onto a galaxy
needs to be inclined to some extent  with respect to the stellar disk
so as to cause the precession of the disk.
The models with smaller $F_{\rm m, gr}$ for a given $f_{\rm v}$
show smaller final ${\theta}_{\rm d}$ owing to weaker tidal perturbation
from infalling groups.
The two models with $f_{\rm v}=0.6$ and $-0.6$ show similar final ${\theta}_{\rm d}$
($\approx 4.5^{\circ}$) yet appreciably
different final ${\phi}_{\rm d}$ (${\phi}_{\rm d}=86^{\circ}$
and $76^{\circ}$, respectively), which indicates that the direction of the orbital angular
momentum of a group infalling onto a galaxy is also important for the final orientation
of the rotation axis of the disk.

The present models with small $F_{\rm m, gr}$ ($<0.01$)  do
not show any significant changes in physical properties of the stellar disk
during and after group infall, if the groups are accreted on the disk galaxy
from outside the virial radius of the galaxy (thus with large $R_{\rm p, gr}$).
Such small groups may well be able to influence the stellar disk if
$R_{\rm p, gr}$ is as small as the disk size of the galaxy ($R_{\rm d}$).
We have investigated the models with $F_{\rm m, gr} <0.01$,
$R_{\rm i}=3R_{\rm d}$, and $f_{\rm v}=0.5$ in order to discuss whether
some of the Galactic satellite galaxies close to the Galactic disk
can cause the Galactic precession and pole shift. In these models,
the most of the dark matter halo of infalling groups can be rapidly
and efficiently stripped by the strong tidal field of the disk galaxy.
Therefore, the central dwarf galaxies  embedded in the original cores
of dark matter halos of their host groups can influence the stellar disk
in these models.

Fig. B4 shows that although more massive groups
can reach closer to the stellar disk owing to more effective dynamical
friction,
the differences in $R_{\rm p, gr}$ between the three models are not so
remarkable.  As a result of $R_{\rm p, gr}$ being smaller than
$R_{\rm d}$,  the galaxy-group (or galaxy-dwarf) interaction can dynamically
influences the stellar disk.  Fig. B5 shows that if $F_{\rm m, gr} \ge 0.003$,
then the final ${\theta}_{\rm d}$ can be larger than $4^{\circ}$ (which
is as large as that derived in the standard model with
much larger $F_{\rm m, gr}$). The final  ${\theta}_{\rm d}$
is larger for larger $F_{\rm m, gr}$ owing to stronger tidal perturbation
of galaxy-group interaction and this dependence can be seen in other models
with different $f_{\rm v}$ and ${\theta}_{\rm d}$.
It should be stressed that although the model with $F_{\rm m, gr}=0.009$ shows
a large ${\theta}_{\rm d}$ at $T=5.6$ Gyr,  the central dwarf galaxy
has not yet merged
with the disk galaxy.

Recent dynamical models of Sgr that can explain the observed physical properties
of the stellar streams associated with Sgr have shown that
$M_{\rm Sgr}=6.4^{+3.6}_{-2.4} \times 10^8 {\rm M}_{\odot}$
and $R_{\rm p, Sgr} \approx 15$ kpc (See Figure 7 in Law \& Majewski 2010).
Therefore, it seems that the original bound stellar mass estimated by
Law \& Majewski (2010) is well below the required mass of Sgr that can
cause the Galactic precession. However the above mass estimation
does not include the total mass of the dark matter halo which
Sgr initially had when Sgr first passed by  the Galaxy. If Sgr initially
had the dark mater halo with the total mass being more than ten times
larger than  that of the baryonic component,  then $M_{\rm Sgr}$ can be much
larger than the required mass ($=3 \times 10^9 M_{\odot}$) for the Galactic
precession. Thus, although the present Sgr is unlikely to cause the
precession, it might have caused the pole shift
when it first passed by  the Galaxy.

\section{The distribution of stripped LMC halo stars}

\begin{figure}
\psfig{file=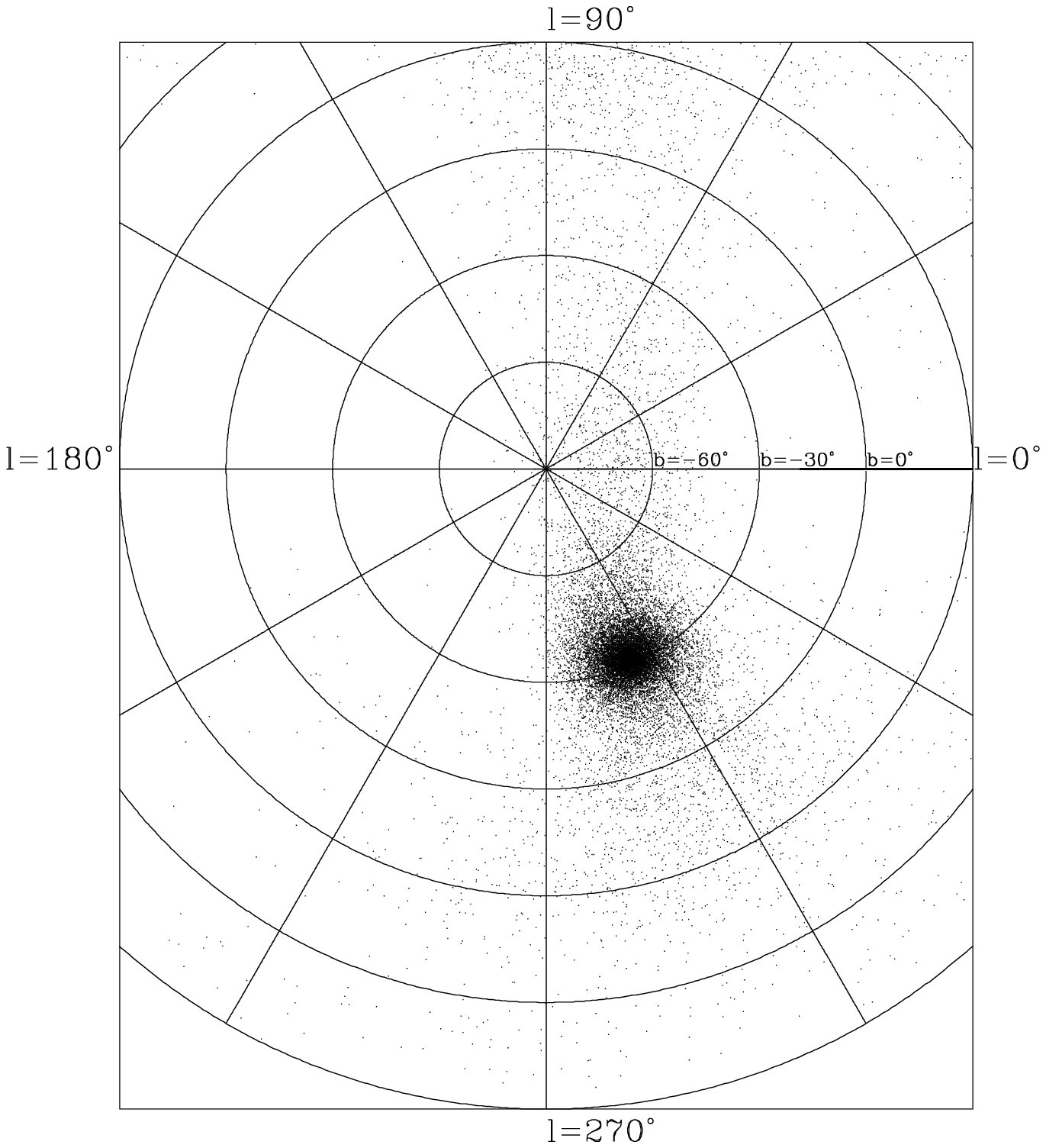,width=8.0cm}
\caption{
The distribution of the stripped LMC stellar halo stars 
in the Galactic coordinate for the standard model. 
}
\label{Figure. 25}
\end{figure}

Bekki (2011) has recently investigated the distribution of stars stripped
from the LMC stellar halo
in the Galactic halo and suggested that the structure and
kinematics of the stripped LMC stars can have fossil information on
the epoch of the LMC accretion onto the Galaxy. However, the present 
position of the LMC was not reproduced well by Bekki (2011). 
Therefore the  simulated 3D distribution of the stripped LMC halo stars can not
be directly compared with the observed one  that will be derived by future 
observational studies. 
We thus adopt  the same LMC stellar halo model as that in Bekki (2011) and thereby
investigate the distribution of the stripped LMC halo stars in the Galaxy.

Fig. C1 shows the distribution of the stripped
LMC halo stars projected onto the Galactic coordinate ($l$, $b$) in the
standard model T1. Although the stripped stars are located in the region
where the Magellanic Stream exists,  they do not form a narrow
tidal stream  there. The LMC halo extends to the SMC region and seems
to be like a common halo that includes the LMC and the SMC.
The predicted physical properties (e.g., locations and radial
velocities)  of the stripped LMC halo stars can be used for 
detecting the stars by SkyMapper telescope (Keller et al. 2007).

\end{document}